\documentclass[a4paper,11pt]{article}
\usepackage{jinstpub} 
\usepackage{lineno}
\graphicspath{{figures/}}
\usepackage{gensymb}
\usepackage{placeins}

\title{\boldmath Characterisation of Thick Gaseous Electron Multipliers  as charge readout operated in pure argon}

 \author[a,1]{G. Eurin, \note{Corresponding author.}}
 \author[a]{A. Delbart,}
 \author[b]{R. De Oliveira,}
 \author[b]{A. Drozd,}
 \author[a,b]{P. Granger,}
 \author[a]{M. Karolak,}
 \author[b]{B. Mehl}

\affiliation[a]{IRFU, CEA, Université Paris-Saclay, F-91191 Gif-sur-Yvette, France}
\affiliation[b]{CERN, The European Organization for Nuclear Research, 1211 Meyrin, Switzerland}

\emailAdd{guillaume.eurin@cea.fr}

\abstract{
The gain measurements of several 1 mm-thick, 10~$\times$~10~cm$^2$ Thick Gaseous Electron Multipliers (ThGEMs), operated in pure argon at 3.3 bar and room temperature are presented. 
Electrostatic simulations, performed with the COMSOL~MultiPhysics\textregistered~software, were employed to guide the design of the detectors, and the field configurations are discussed.
A modified ThGEM design, incorporating two embedded internal electrodes in addition to the conventional top and bottom electrodes, was developed to mitigate discharge-induced instabilities and to enable operation at higher gains.
Several such structures were produced and compared to the standard 1 mm-thick two-electrode ThGEM. 
Gain measurements were conducted for every design described here over a wide range of applied electric fields and up to large values.
The experimental setup and measurement methodology are described, alongside a comparative analysis of the performance of the different detector geometries.

}

\begin{document}
\definecolor{gray}{gray}{0.75}

\setcounter{tocdepth}{2}
\maketitle
\flushbottom

\section{Introduction}
\label{sec:Introduction}
Micromegas \cite{GIOMATARIS199629} detectors and Gas Electron Multipliers (GEM) \cite{Sauli:1997qp} have revolutionised the field of particle detection and imaging, giving rise to the family of Micro-Pattern Gaseous Detectors (MPGD). 
They lead to the enhancement of the rate capabilities, the sensitivity, and the spatial resolution of various instruments.
One of the devices derived from the original GEM is the Thick Gaseous Electron Multiplier (ThGEM) \cite{CHECHIK2004303}, also known as the Large Electron Multiplier (LEM). ThGEMs differ from the original GEM design by employing a thick structure composed of copper-clad electrodes, separated by an insulating layer, in which large holes are mechanically drilled.
This structural configuration imparts notable robustness and simplifies integration, especially for their use in pure cryogenic noble gas environments. Beyond particle physics, ThGEM technology has found applications in other fields such as medical imaging \cite{Imaging}. 

In this work, the potential of the ThGEM as a readout solution is investigated.
The operation principle and strengths of a ThGEM are presented in Section~\ref{sec:design}.
The experimental setup, the different detector designs studied, and the associated data analysis procedures are described in Section~\ref{sec:technologies}.
The methodology used for gain determination is described in Section \ref{sec:gain}.
Finally, performance measurements and results are compiled in Section~\ref{sec:results}.

\section{Thick GEM Technical Design and Production}
\label{sec:design}

\subsection{Operating principle of a ThGEM}
\label{sec:operation_principle}
\begin{figure}[htbp]
\centering
\includegraphics[width=\textwidth]{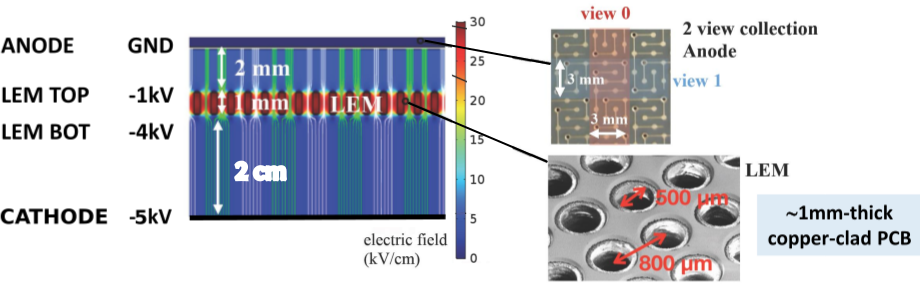}
\caption{Left: Operating electrostatic conditions of the ThGEM. The typical operating potentials are indicated on the left. Top Right: Design of the anode reading out the ThGEM, with the two readout views. Bottom right: Microscopic image of the holes drilled through the ThGEM highlighting their dimensions. 
\label{fig:ThGEM_principle}}
\end{figure}

The main and common design parameters of the ThGEMs studied here are those chosen and used for a Dual-Phase argon Time-Projection Chamber (TPC) readout, illustrated in Figure~\ref{fig:ThGEM_principle}, and described in reference \cite{Cantini_2015}.
The ThGEM consists of a series of holes with a diameter of $\sim$~500~$\upmu$m.
The hexagonal pitch of the holes is defined as $\sim$~800~$\upmu$m. 
The holes are mechanically drilled through the polymer-electrode stack.
The thickness of FR4 (PCB - printed circuit board - base material made from a flame retardant epoxy resin and glass fabric composite) and copper-clad electrodes is $\sim$~1~mm and this value directly affects the amplification capabilities of the ThGEM and therefore its gain. 
All prototypes were produced with the same batch of 1 mm thick raw FR4 material for a mean total measured thickness of the ThGEMs produced of $\sim$1.1~$\pm$~0.03~mm (with copper electrodes).
Depending on the ThGEM design, this core stack can also be a multilayer structure with additional layers such as embedded electrodes within the bulk of the stack, or it may possess an additional layer on top of the electrodes, e.g. a resistive layer. This is further developed in the following sections.

\begin{figure}[htbp]
\centering
\includegraphics[width=0.7\textwidth]{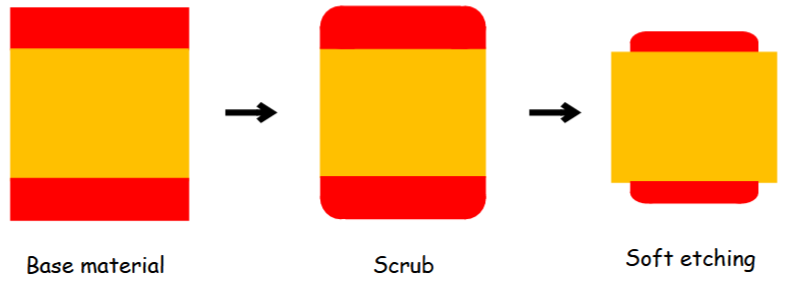}
\caption{Production process using "global etching" for the rims around ThGEMs holes. In yellow, the FR4 base material surrounded by the orange copper-clad electrodes \cite{deoliveira}. On the rightmost sketch, the location of the rims is visible as the area without copper on the edge of the holes.
\label{fig:ThGEM_rims}}
\end{figure}

To ensure the electrical stability of ThGEM, rims around the holes have been produced using the "global etching" process to remove a ring of copper $\sim$~40~$\upmu$m wide  around the holes on each side of ThGEM (see Figure~\ref{fig:ThGEM_rims}).

\subsection{ThGEM operating conditions}
\label{sec:ThGEM_operation}
The ThGEMs were operated under the same electrostatic conditions as the one used for a dual-phase argon TPC \cite{cotte:tel-02382815}.
As can be seen in Figure~\ref{fig:ThGEM_principle}, several volumes can be defined.
The volumes are created by the various elements of the setup: the cathode, the electrodes of the ThGEM and the anode.

The drift volume corresponds to the space between the cathode and the bottom electrode of the ThGEM.
When a charged particle passes through the drift volume, it ionizes the gas producing electron-ion pairs.
These primary electrons are accelerated towards the ThGEM due to the application of an electric drift field of $\sim$~500~V/cm.
The value of the drift field has an impact on the overall electronic transparency of the setup, which will be further described in section~\ref{sec:transparency}.

The next volume is the amplification volume between the two electrodes of the ThGEM. 
A much larger electric field is applied here, with values reaching above 30~kV/cm.
As the electrons traverse this amplification volume, they undergo Townsend avalanche multiplication.
In this process, the high electric field causes electrons to gain enough kinetic energy to ionize additional gas molecules. 
This produces secondary electrons, which are then in turn accelerated, leading to a cascading effect. 
Each stage of the avalanche contributes to the overall amplification, resulting in a significant increase in the number of electrons compared to the primary electrons initially produced by ionisation.
The amplification field is therefore governing the value of this multiplication factor, defined as the gain of the ThGEM ($G_{ThGEM}$), which is the variable that is studied in this work.
The goal is to increase the field to the highest accessible value while maintaining a stable operating detector.

The final volume is the collection volume between the anode and the top electrode of the ThGEM.
No further amplification is expected here.
However, thanks to a collection field with values around 5~kV/cm, the secondary electrons are collected onto the anode.
This produces an analogue electrical signal that is read out by the anode and the associated electronics. The anode is the one described in \cite{Cantini_2014}. Its two-dimensional (x-y) copper strips pattern, with a 3~mm pitch, is designed to lower the capacitance load when connected to the readout electronics ($\sim$~140~pF/m) and to enable a track localisation with a spatial resolution of the order of the millimetre (see picture on Figure ~\ref{fig:ThGEM_principle}).
The collection field is defined in order to maximise the collection efficiency onto the anode which is assumed here to be 100~\% \cite{cotte:tel-02382815}.

\subsection{ThGEM production}
The ThGEM prototypes were produced by the CERN/MPT workshop which specialises PCB technologies and the specific process required to produce ThGEMs. 

One of the main challenges is the drilling of the holes.
Given the large number of holes that have to be produced, the machine must automatically exchange the drill due to wear effects.
It is estimated that the drill needs to be replaced approximately every 1000 holes.
This is critical, as the quality of the holes governs the stability of the ThGEM.
It is even more crucial for the embedded electrodes design (see section \ref{sec:embedded_electrodes}) since the inner electrodes must be inside a layer of insulating material in every hole at the risk of having a non-functioning detector.

The next most important aspect to be mastered to ensure the stability of the ThGEM is the quality of the rims (see section~\ref{sec:operation_principle}).
The global etching process was used to produce the rims (see Figure \ref{fig:ThGEM_rims}), removing copper uniformly around the edges of the holes and reducing the external electrodes thickness.

The ThGEMs with two embedded electrodes were produced with state of the art multilayer PCB manufacturing processes.
The core is made of 0.4 mm or 0.6 mm double-sided copper clad FR4 sheet for EE04 ThGEM or EE06 ThGEM respectively.
This core is copper etched with masks to make the 2 internal electrodes, and the required thickness of insulator and copper layers are then pressed on both sides to reach 1~mm in thickness. 
The holes are drilled in this 1~mm thick stack and the rims are made with the same global etching process as for standard ThGEMs, but with an additional critical step for the alignment of the drilling machine with the internal electrodes.

\subsection{Sparking rate and maximum operating voltage}
The ThGEMs being electrostatic detectors under high electric fields, they are subject to sparking.
This corresponds to the release of a local build-up of charges leading to the formation of an electrical arc within a hole and creating a connection between the two outer electrodes.
The consequence of these sparks is the release of the accumulated charges leading to the drop of the high voltage, and the associated dead time needed to recover the gain in the detector. 
The duration of this dead time is determined mainly by the specifications of the high-voltage power supply and the capacitance of the connected electrodes. 
In our setup, for 10x10~cm$^2$ ThGEMs, this dead time was measured to be around 5~s. 
The maximum operational voltage V$_{\text{max}}$ at which we operated our ThGEMs was determined by a maximum sparking rate of four sparks per minute, corresponding to 30\% of operation dead time. 

\section{Explored ThGEM Designs}
\label{sec:technologies}

\subsection{Context of the large scale production of 50x50~cm$^2$ ThGEM for a Dual-Phase TPC}
\label{sec:initial}

\begin{figure}[htbp]
\centering
\includegraphics[width=\textwidth]{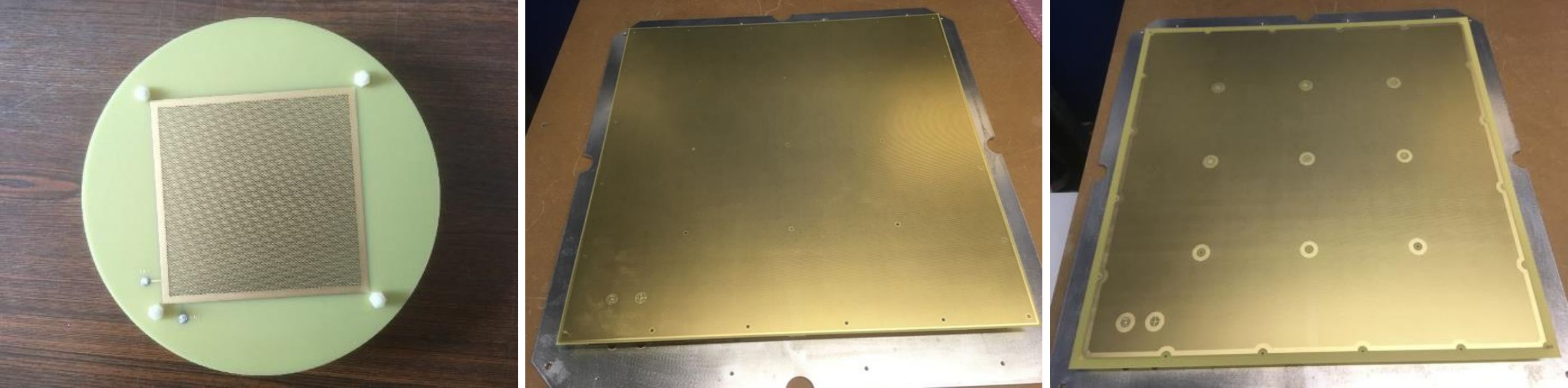}
\caption{Pictures of the various versions of ThGEMs developed at IRFU. Left: 10$\times$10~cm$^2$ prototype, Middle: CFR-34 50$\times$50~cm$^2$ detector, Right: CFR-35 50$\times$50~cm$^2$ detector.
\label{fig:lems}}
\end{figure}

\begin{figure}[htbp]
\centering
\includegraphics[width=0.6\textwidth]{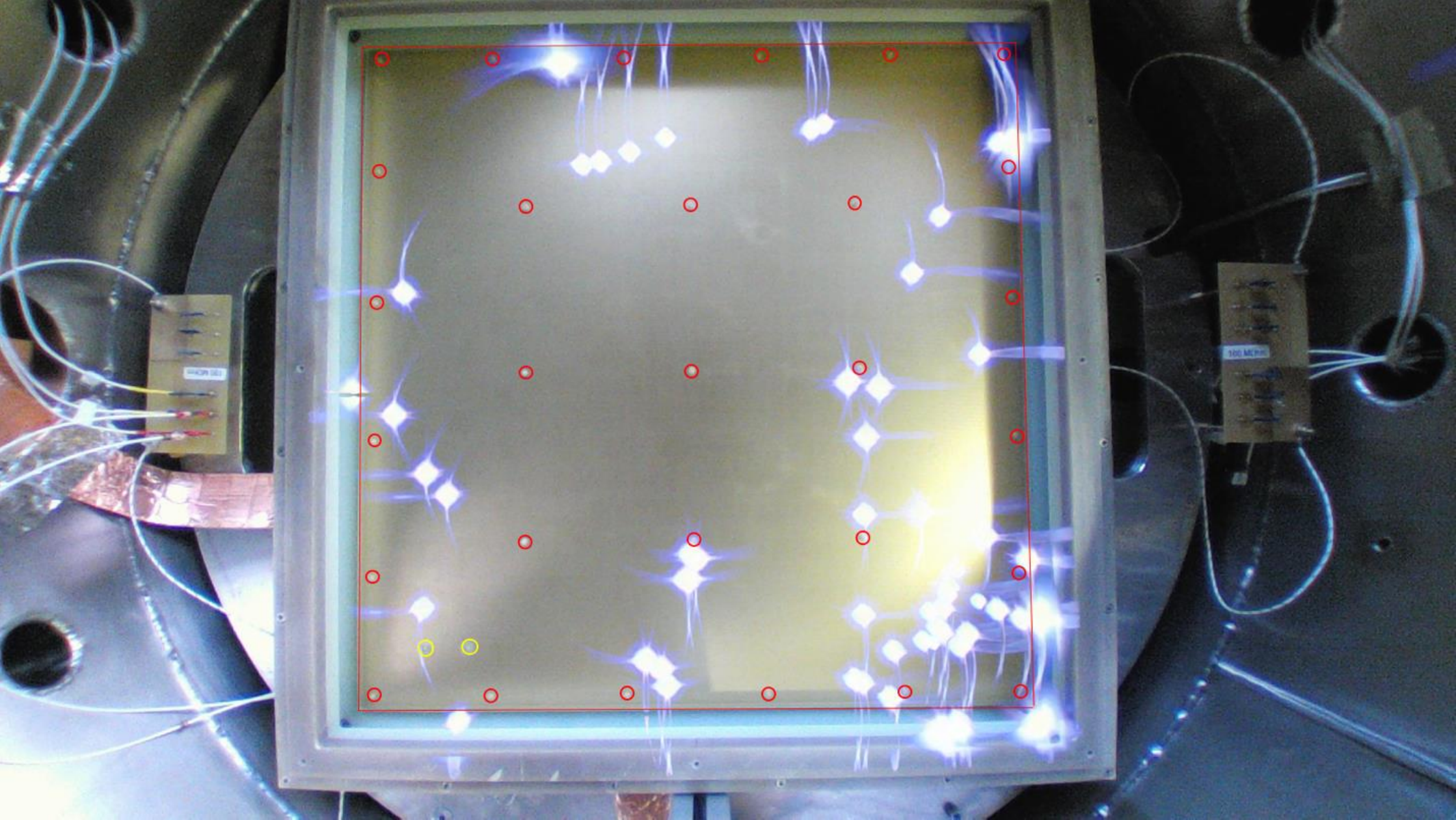}
\caption{Picture of the sparks observed in the pressurised vessel on a CFR-34 ThGEM taken by a camera. The image is obtained by stacking several images.
\label{fig:sparking}}
\end{figure}

Prior to this study, during the development of ThGEM at IRFU aiming at the production of 50$\times$50~cm$^2$ ThGEMs, several electrostatic issues occurred.
This led to the redesign of the ThGEMs to mitigate these issues and achieve stable operational conditions.
As illustrated in Figure~\ref{fig:lems}, 10$\times$10~cm$^2$ prototypes were initially produced in order to test the manufacturing process prior to the production of large ThGEMs.
The CFR-34 design was the 50$\times$50~cm$^2$ upscaling of this prototype which turned out to present significant sparking, mainly around the edges of the detector, as illustrated in Figure~\ref{fig:sparking}.

\begin{figure}[htbp]
\centering
\includegraphics[width=\textwidth]{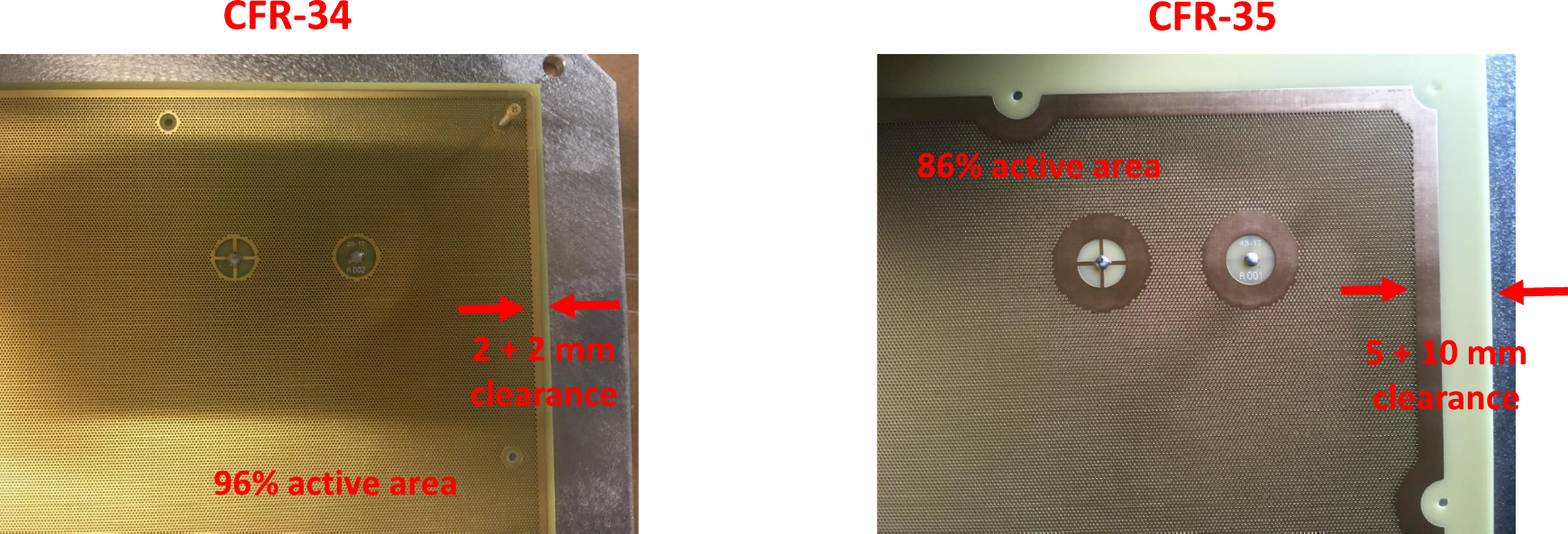}
\caption{Illustration of the main differences between the CFR-34 and CFR-35 ThGEMs.
\label{fig:cfr34_35}}
\end{figure}

In order to mitigate this sparking issue, a new design was conceived.
The CFR-35 concept consisted in the reduction of the total active area from 96~\% to 85~\% as visible in Figure~\ref{fig:cfr34_35}.
This translated into the removal of the copper in the areas most prone to sparking, namely around the mechanical assembly holes and on the edges of the detectors.

The actions taken to limit the instabilities were successful.
The initial target of 1 spark per detector per 36~h while keeping a gain of several units could be achieved \cite{grang2022}.
Despite this, novel ideas were further investigated.
They allowed larger gains to be reached without sacrificing the active area.

The standard design refers in the following to the design with two electrodes surrounding an insulator layer, as for the CFR-34 and CFR-35 detectors.
Several technologies have been explored to go beyond the requirements of ThGEMs for their initial purpose.

\subsection{ThGEM designs explored}
\label{sec:embedded_electrodes}
\paragraph{ThGEM with connected embedded electrodes}
One of the most promising designs corresponds to the embedded electrodes design.
One of the main limitations to the stability and possibly to the effective gain of the ThGEMs is coming from the electric field across the two electrodes.
Indeed, the larger the electric field, the more electrostatic phenomena such as plasma and electrical arc creation can occur.

\begin{figure}[h!]
\centering
\includegraphics[width=0.6\textwidth]{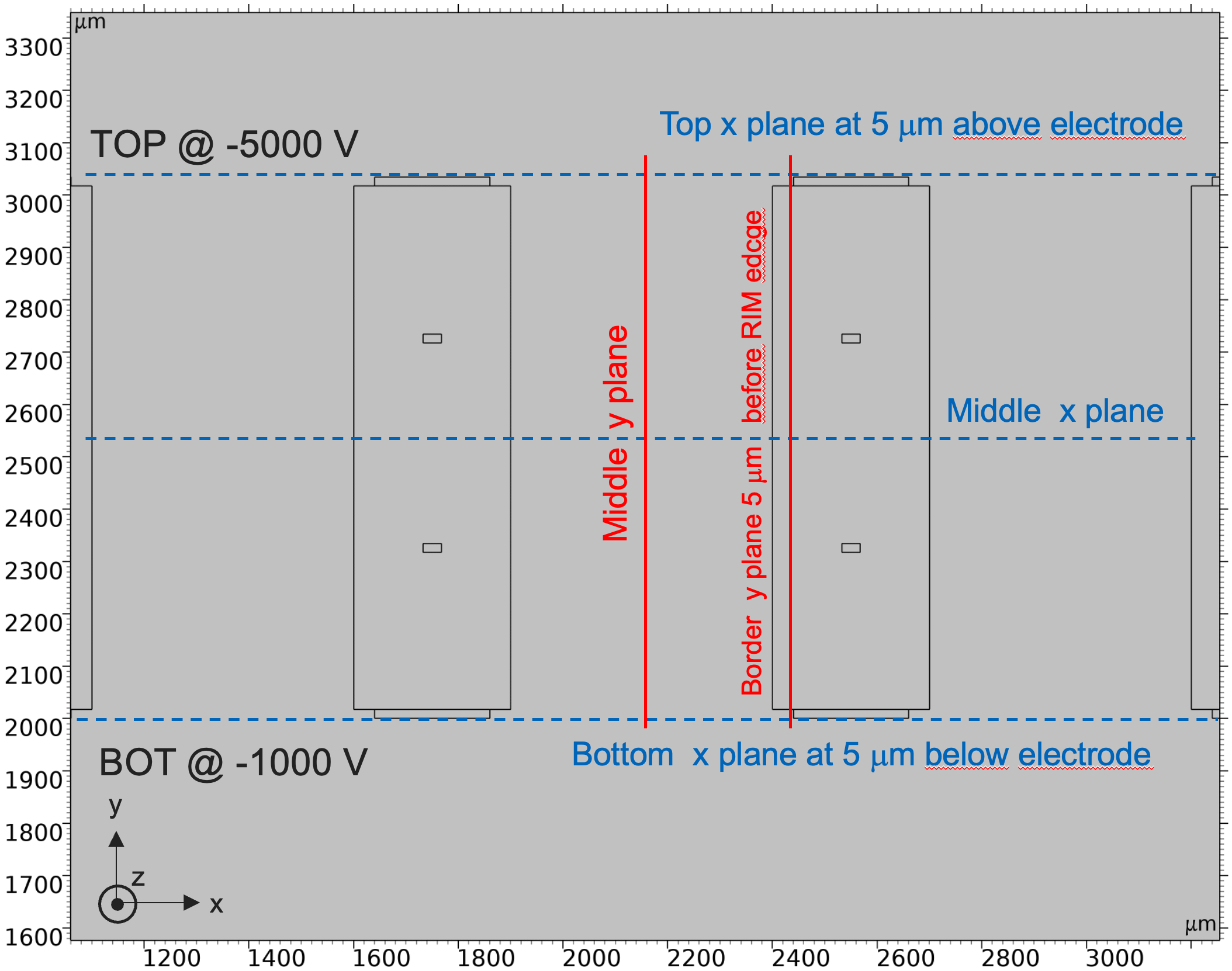}
\caption{COMSOL\textregistered~simulation setup of a ThGEM with embedded electrodes (EE04). The grounded anode (not shown) is at y=0, and the cathode drift plane, set to -6000 V, is 2 cm above the TOP electrode. Measurement planes are marked in red for vertical planes and in blue for horizontal planes.
\label{fig:buried_electrodes}}
\end{figure}

In order to reduce the electrostatic instabilities, one solution is to repel the field lines away from the edges and the inner surfaces of the holes.
Indeed, any inhomogeneity in the material in these areas could lead to strong instabilities.
This can be achieved by adding extra electrodes inside the bulk of the PCBs themselves.
These electrodes are then polarised to the same voltage as the closest electrode.
In this case, two additional electrodes are placed within the ThGEM stack in order to shape the electric field.
Two 10$\times$10~cm$^2$ ThGEM prototypes were produced and tested following the electric field simulations with COMSOL~MultiPhysics\textregistered~\cite{comsol}~described in section \ref{sec:comsol}, labelled EE04 and EE06 (see Figure \ref{fig:buried_electrodes}). 
The design parameters of these two prototypes are the same as the CFR-34 or CFR-35 ThGEM (overall thickness, rim size, hole diameter, hexagonal holes pattern with same pitch between holes) but with two 35 microns wide internal electrodes placed at 0.4~mm or 0.6~mm from each other symmetrically with respect to the middle x-plane (see Figure~\ref{fig:buried_electrodes}) for the EE04 or EE06 respectively. 
A large FR4 border surrounds the active area in order to prevent instabilities due to border effects at the edges of the active area (see picture of a 10x10 cm$^2$ ThGEM in Figure \ref{fig:lems}).

\paragraph{ThGEM with independent embedded electrodes}
The embedded electrode design was further explored.
Instead of keeping the same potential on the outer electrodes and on the closest respective embedded electrodes, each electrode is left independent.
This entails the creation of 4~HV connections connected to individual power supply channels.
The electric field can therefore be set into an \textit{inverse} configuration, further described in section~\ref{sec:comsol}.

\noindent Two 10$\times$10~cm$^2$ ThGEM prototypes were produced and tested following the same protocol as the detectors described in the previous section, labelled EE04I and EE06I.
The results obtained with these detectors are described in section \ref{sec:gain_results}.

\subsubsection{COMSOL\textregistered~ simulations of the electric field}
\label{sec:comsol}
The aim of the embedded electrodes is to tune and shape the electric field.
This should lead to an improvement of the ThGEM stability by reducing high electric field "hot" spots at the edges of the holes.
This is achieved while maintaining a sufficiently high electric field inside the ThGEM holes, which is necessary to trigger secondary electron production via the avalanche process. 
2D Electric field lines and Electric field magnitude were simulated with the COMSOL~MultiPhysics\textregistered~finite element simulation software for the experimental setup configuration used for the ThGEM qualification described in section \ref{sec:ThGEM_qualification} and a fixed potential difference across the ThGEM external electrodes of 4000 V. 
The COMSOL\textregistered~ model is described in Figure \ref{fig:buried_electrodes}.\\

\begin{figure}[h!]
\centering
\includegraphics[width=0.6\textwidth]{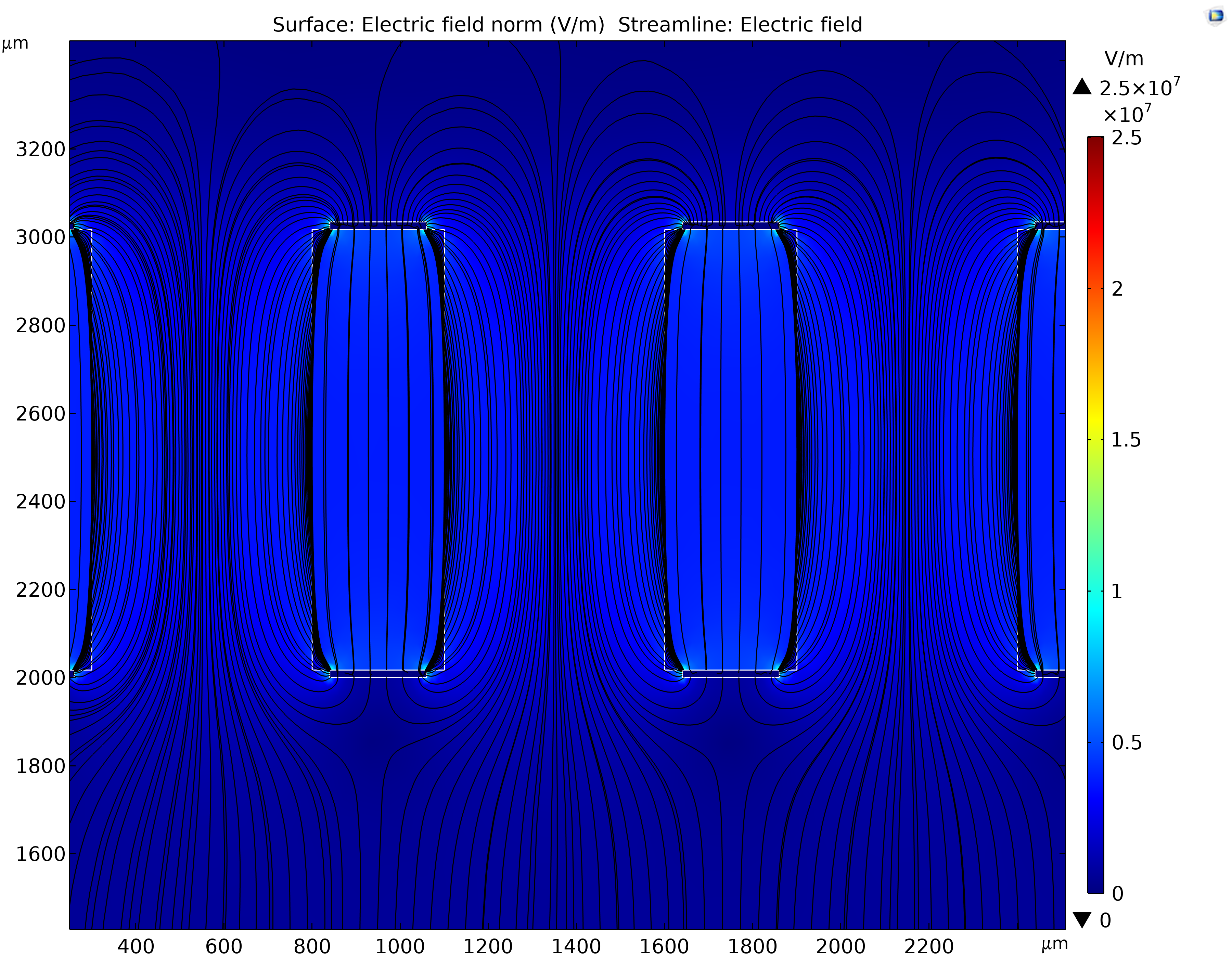}
\includegraphics[width=0.6\textwidth]{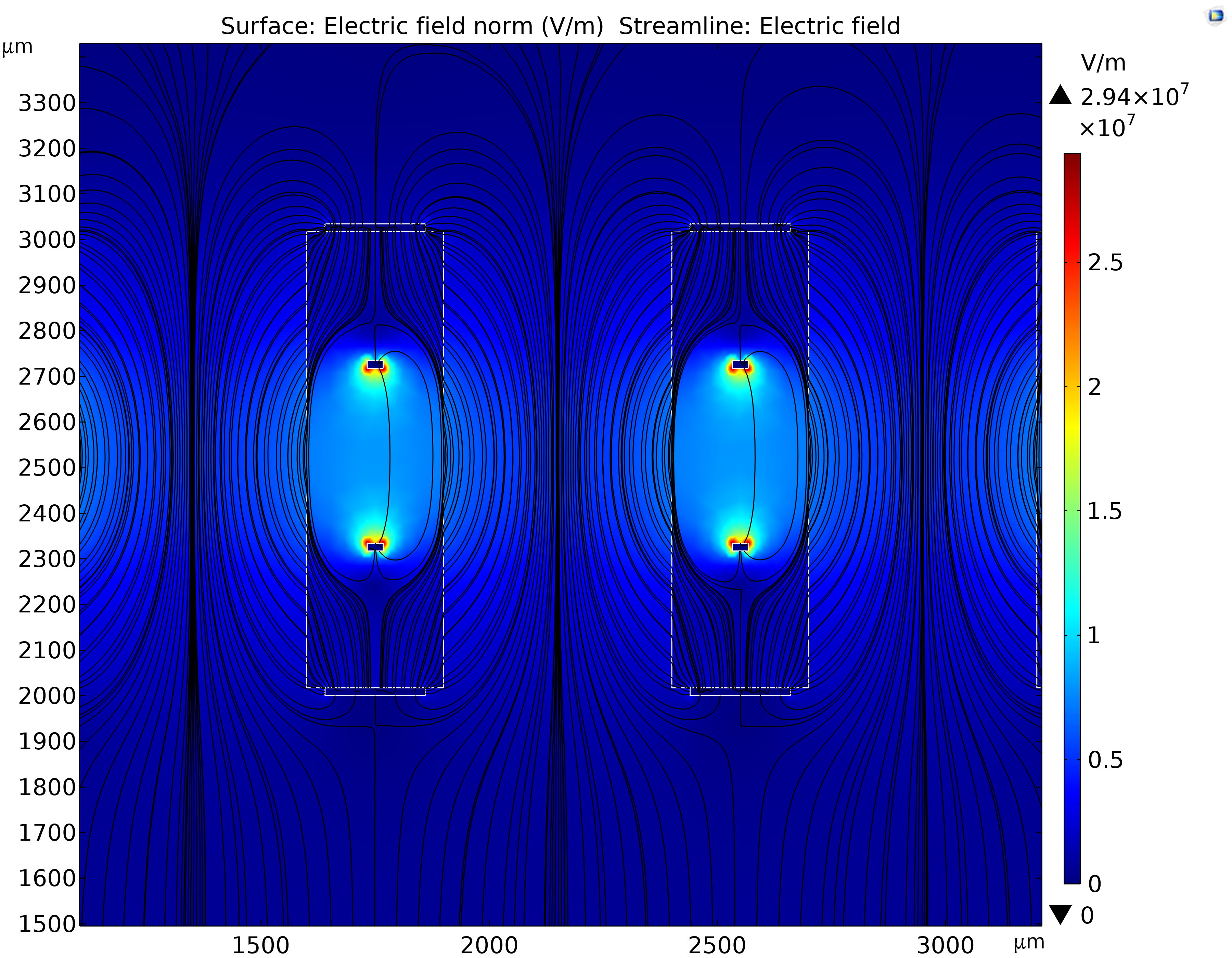}
\includegraphics[width=0.6\textwidth]{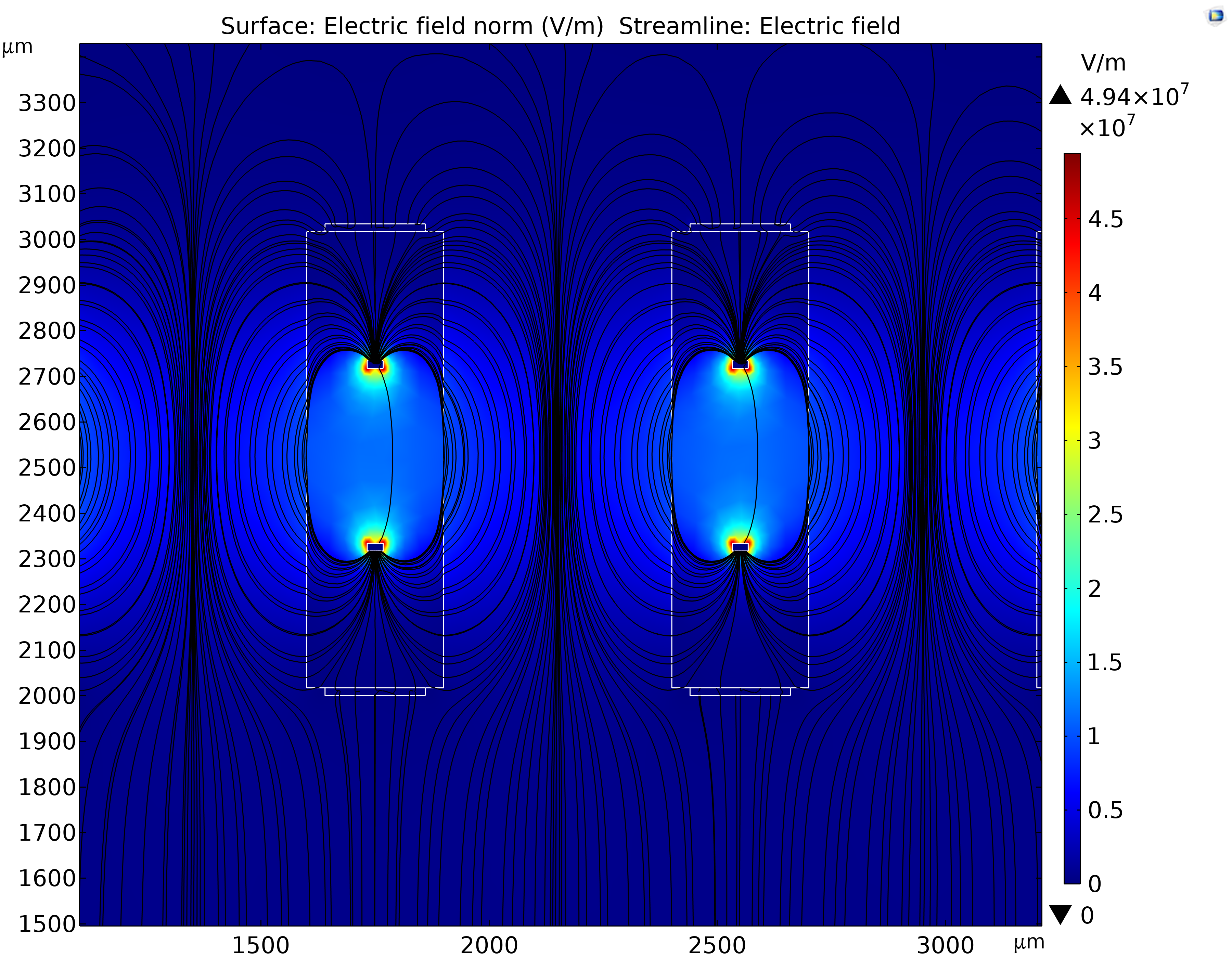}
\caption{COMSOL\textregistered~ simulation of the electric field streamlines in x-y plane for three ThGEM configurations (see text and Figure~\ref{fig:buried_electrodes} for details). Top: reference ThGEM without embedded electrodes, Middle: ThGEM with embedded electrodes at the same potential as the external electrodes, Bottom: ThGEM with embedded electrodes in a 1000V inverse field configuration.\label{fig:field_lines}}
\end{figure}

\begin{figure}[h!]
\centering
\includegraphics[width=0.6\textwidth]{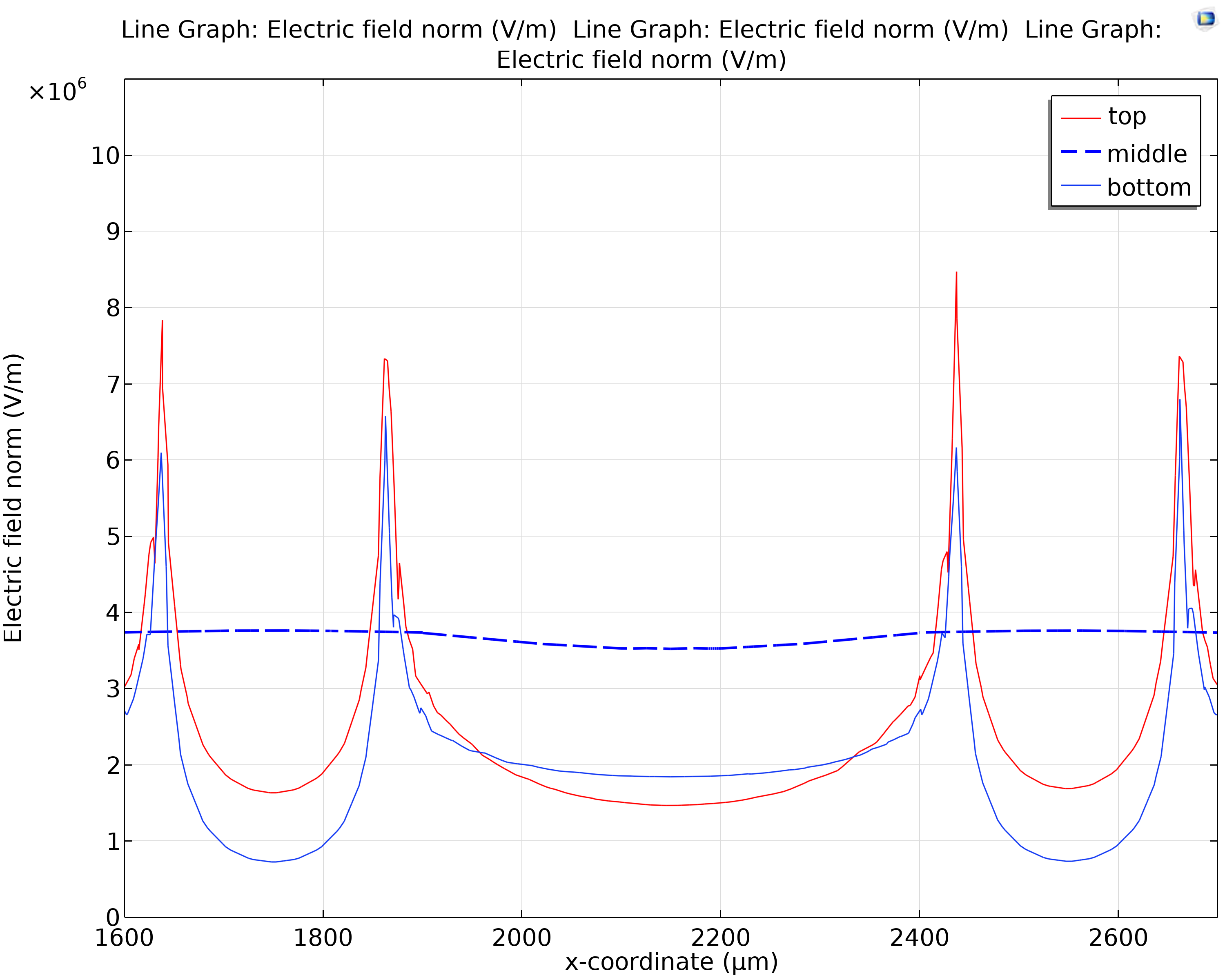}
\includegraphics[width=0.6\textwidth]{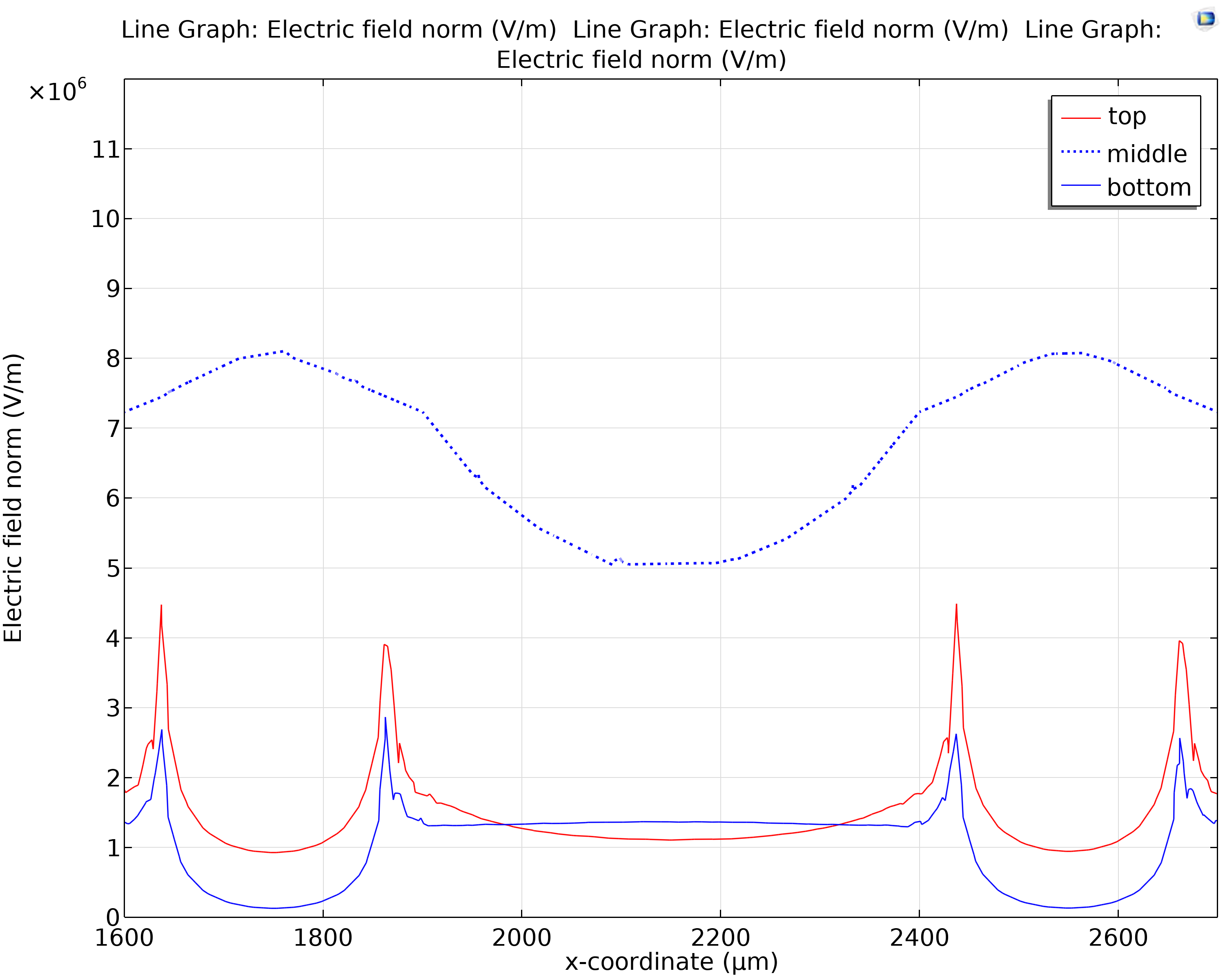}
\includegraphics[width=0.6\textwidth]{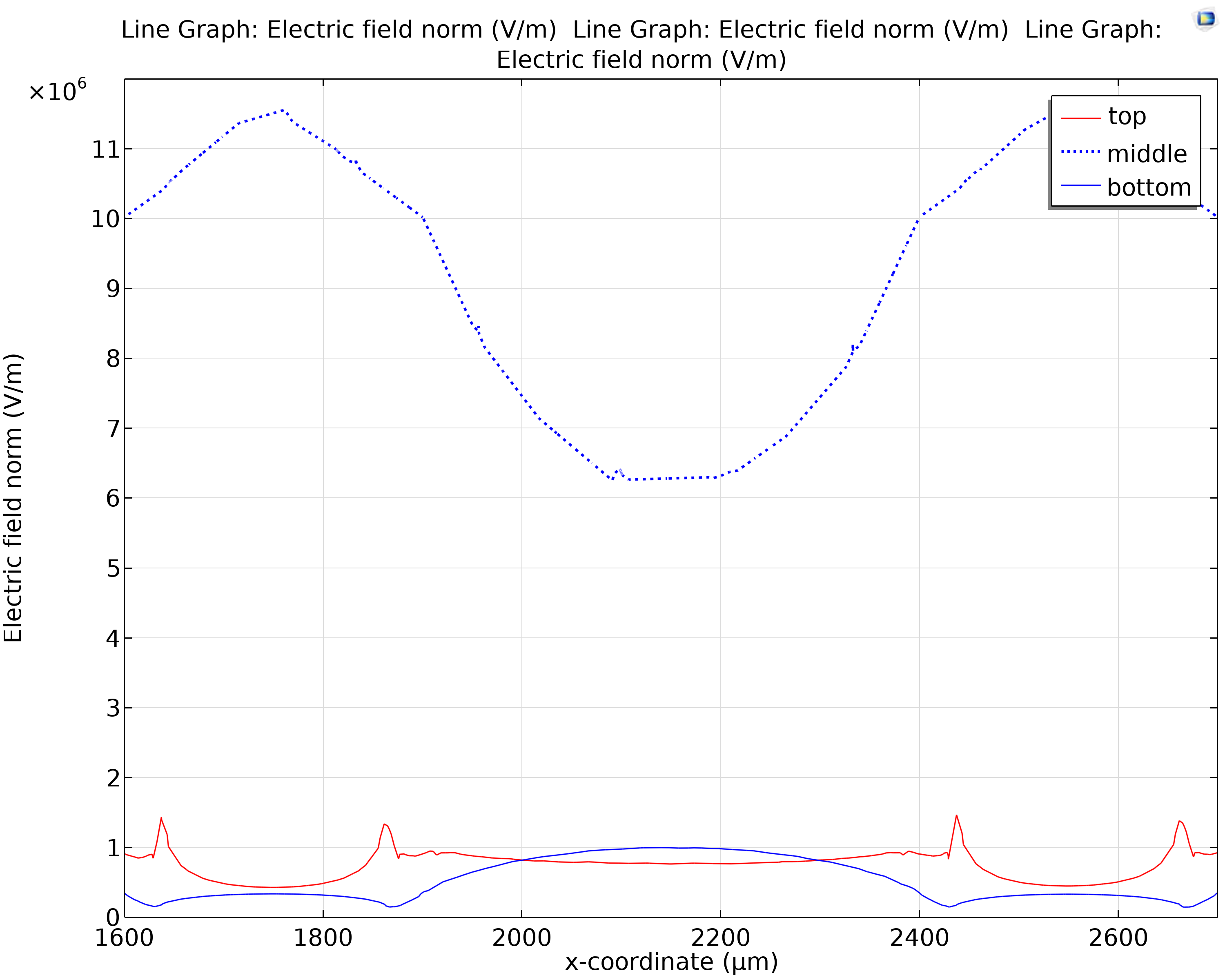}
\caption{COMSOL\textregistered~ simulation of the electric field magnitude in x planes for the three ThGEM configurations (see text and Figure~\ref{fig:buried_electrodes} for details). Top: reference 1 mm ThGEM without embedded electrodes, Middle: ThGEM with embedded electrodes at the same potential as the external electrodes, Bottom: ThGEM with embedded electrodes in a 1000V inverse field configuration. \label{fig:field_simulations_x}}
\end{figure}

\begin{figure}[h!]
\centering
\includegraphics[width=0.6\textwidth]{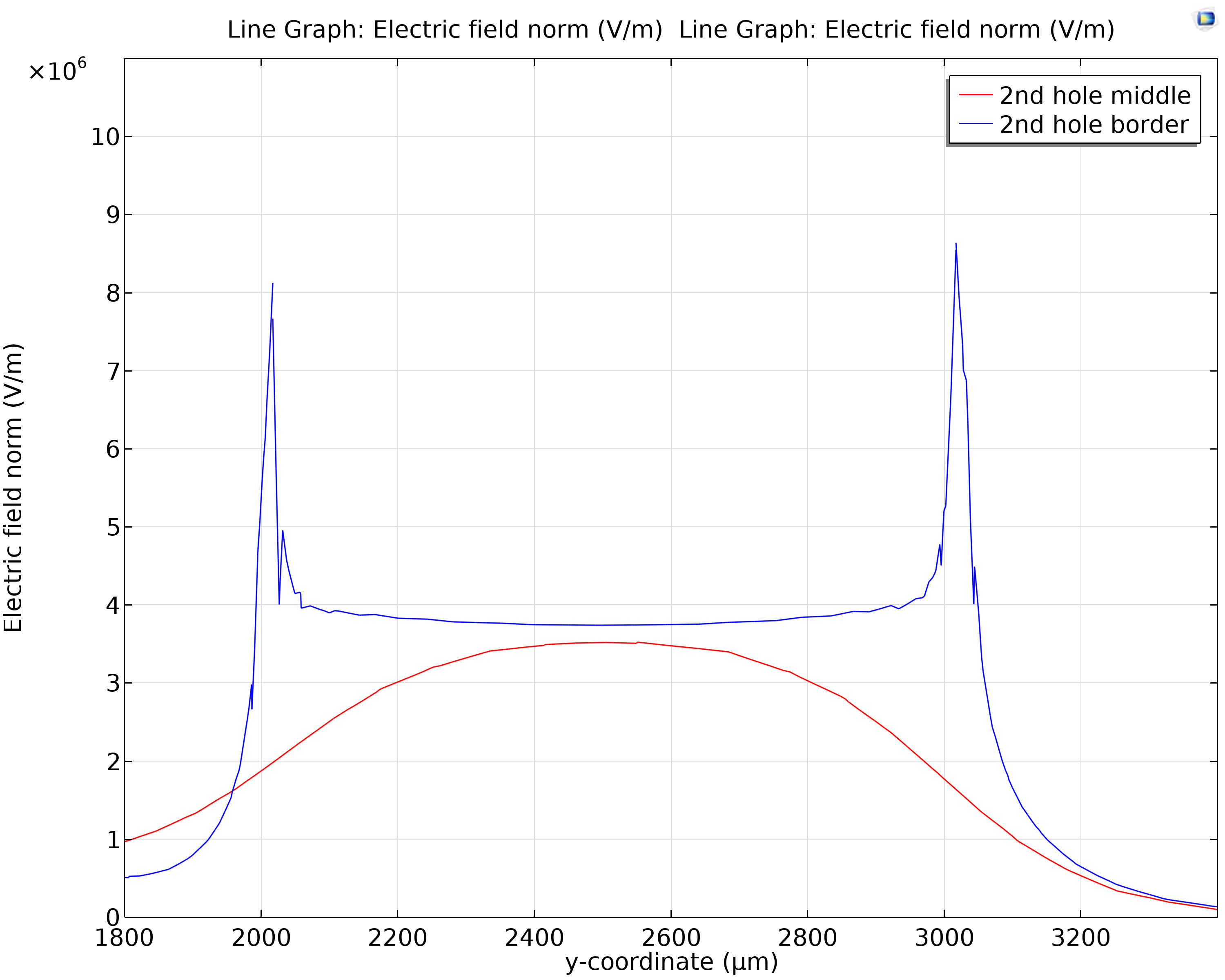}
\includegraphics[width=0.6\textwidth]{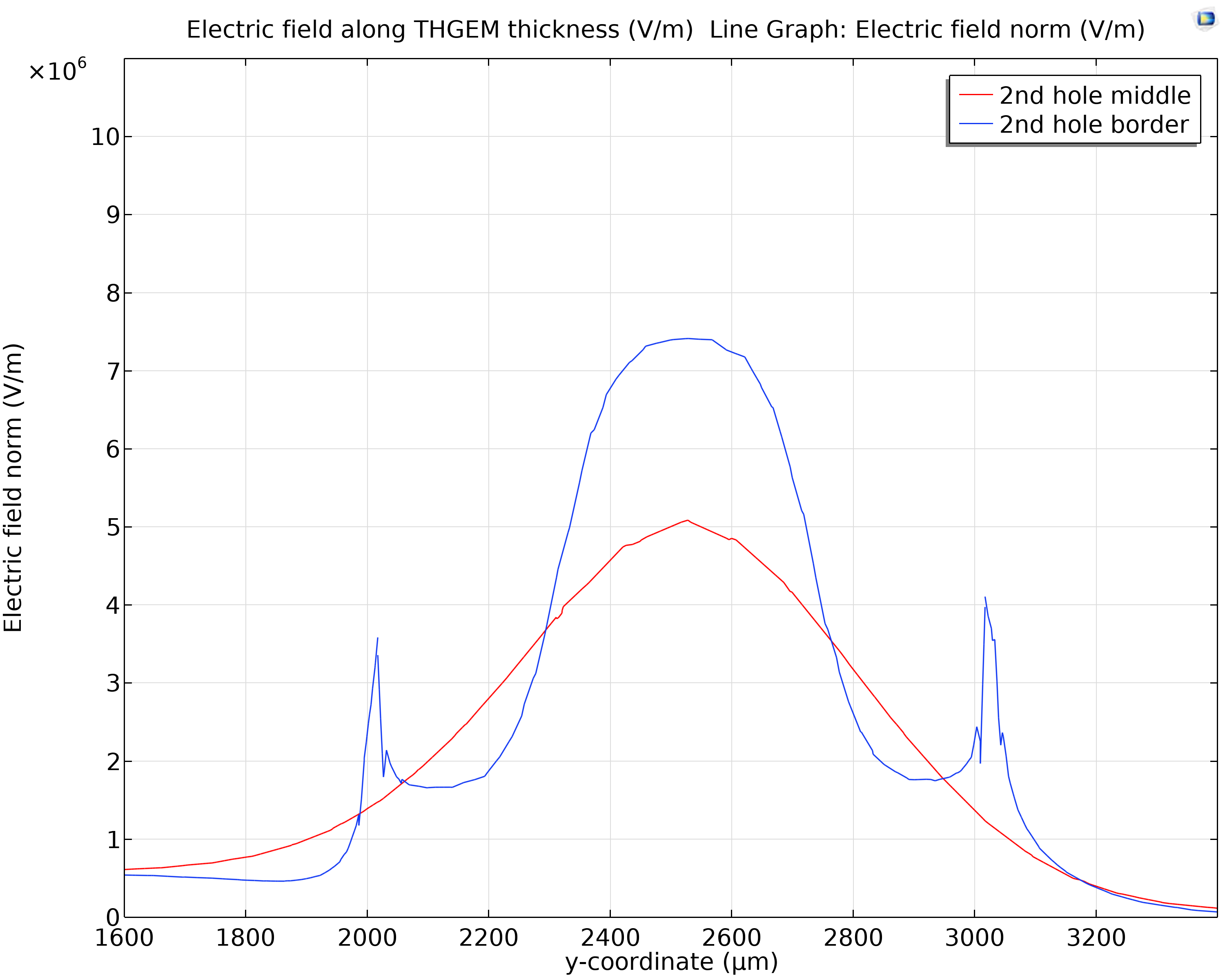}
\includegraphics[width=0.6\textwidth]{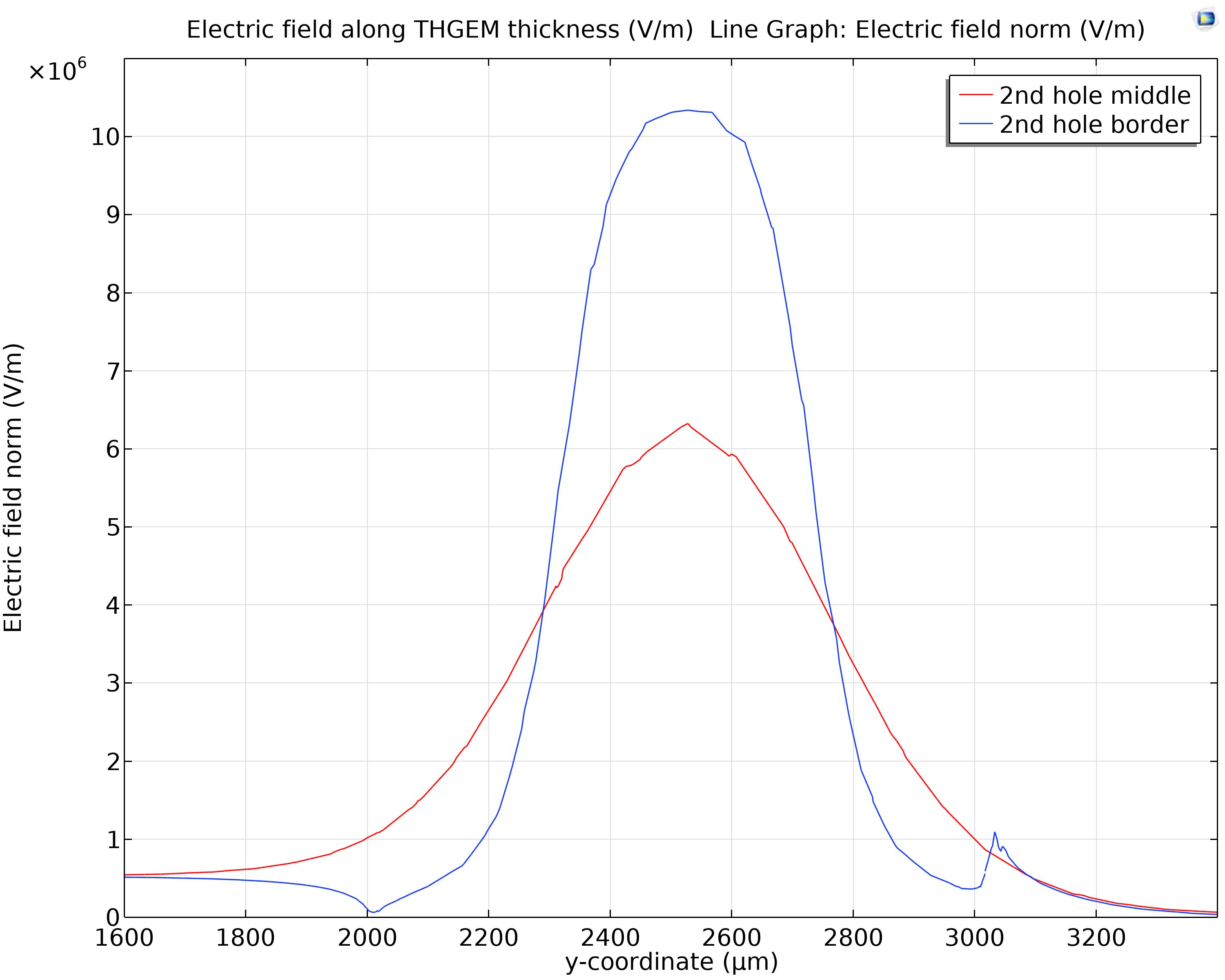}
\caption{COMSOL\textregistered~ simulation of the electric field magnitude in y planes for the three ThGEM configurations (see text and Figure~\ref{fig:buried_electrodes} for details). Top: reference 1 mm ThGEM without embedded electrodes, Middle: ThGEM with embedded electrodes at the same potential as the external electrodes, Bottom: ThGEM with embedded electrodes in a 1000V inverse field configuration. \label{fig:field_simulations_y}}
\end{figure}

Three configurations of ThGEM operated with 4000 V across external electrodes in the configuration of Figure \ref{fig:buried_electrodes} have been simulated :
\begin{itemize}
    \item the standard 1~mm thick configuration of the ThGEM without embedded electrodes,
    \item the EE04I embedded electrodes configuration with the potential of the Top (Bottom) embedded electrode set at the same potential as the TOP (Bottom) external electrode,
    \item the EE04I embedded electrodes configuration with a 1000 V \textit{inverse} electric field between the Top (Bottom) external electrode and the Top (Bottom) embedded electrode, i.e. the Top embedded electrode set at -6000 V and the Bottom embedded electrode set at 0 V.
\end{itemize}
The results of these simulations are illustrated on Figure~\ref{fig:field_lines} for the 2D Electric field lines and on Figures~\ref{fig:field_simulations_x} and~\ref{fig:field_simulations_y} for the electric field magnitude along x and y axes.
Figures~\ref{fig:field_simulations_x} shows the electric field magnitude along the x axis (x-z ThGEM planes) for three planes  illustrated in~\ref{fig:buried_electrodes}, located at y=5~$\upmu$m above the ThGEM TOP electrode (labelled "TOP"), the middle of the ThGEM thickness (labelled "middle") and at y=5~$\upmu$m below the ThGEM bottom electrode.
Figures~\ref{fig:field_simulations_y} shows the electric field magnitude along the y axis (y-z planes orthogonal to the ThGEM plane) for two planes illustrated in~\ref{fig:buried_electrodes}, located at 5~$\upmu$m before the rim edge (labelled "2nd hole border") and in the middle of the second hole of the ThGEM (labelled "2nd hole middle").

Compared to the standard 1~mm thick design, the impact of the use of the embedded electrodes is a strongly reduced electric field on the border of the holes and an increase of the maximum electric field in the middle of the hole.
The 1000 V inverse electric field configuration appears to be the best one regarding these electrostatic criteria.
For a given high voltage difference between external electrodes, this should lead to reduced instabilities with a lower sparking rate and operation dead time. 
One could therefore expect to be able to operate the ThGEM at higher potential difference across the ThGEM external electrodes to reach higher gains for the same conditions of instabilities, i.e. the same operation dead time.

\FloatBarrier

\subsection{ThGEMs qualification conditions}
\label{sec:ThGEM_qualification}

One of the key parameters for MPGDs is the gaseous detection medium, and mainly its density.
Indeed, the gain of the ThGEMs is assumed to be directly linked to the gas density.
In the context of this work, the ThGEMs were designed to operate within an ultra-pure gaseous argon at temperatures close to the liquid argon temperature of 87.3~K.
In order to qualify the detectors without necessitating cryogenic conditions, an equivalent operational gas density was used at room temperature.

This gas density is the one used in the vapour phase of dual-phase argon applications detectors and therefore provides a good estimate of the gain expected under these cryogenic conditions. 
Using the noble gas law, the liquid argon conditions (87.3~K, 1~bar) then lead to an operating condition of 3.36~bar at 20~\degree~C in order to keep a constant gas density.
These are the density conditions that have been applied for the results presented in this article.

During the characterisation step, carbonisation could be observed.
This points to a weak region in the ThGEM from an electrostatic point of view.
This carbonisation corresponds to part of the insulator heating up in presence of a plasma from an electrical discharge.
The affected ThGEMs were treated with KMnO$_4$ at the CERN EP-DT-EF to remove the carbonised FR4 from the hole while leaving the copper surface unaltered.
These ThGEMs were subsequently cleaned at CEA-IRFU and could be further operated.

\subsection{The experimental setup}
\begin{figure}[htbp]
\centering
\includegraphics[width=\textwidth]{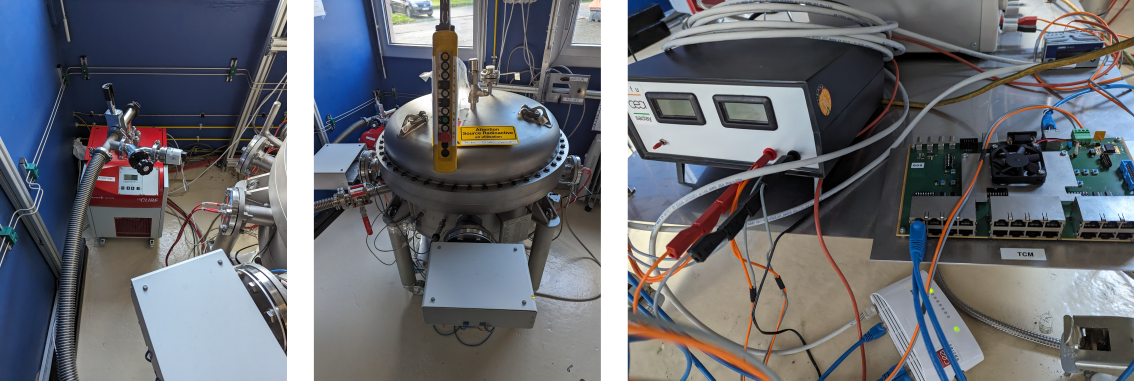}
\caption{Pictures of the pressurised vessel used for the ThGEMs characterisation. Left: pumping station and gas system, middle: the vessel in the laboratory at CEA, right: the electronics power supply and TCM (Trigger Clock Module). 
\label{fig:ESP}}
\end{figure}

In order to perform the qualification of the ThGEMs, a dedicated test setup has been assembled at IRFU as visible on Figure~\ref{fig:ESP}.
It consists of a pressure vessel able to reach an overpressure of 4~bar with the equipment to operate the ThGEMs.
This comprises the gas system necessary to supply the operating gas, to evacuate the vessel and to perform the gas purification (see section~\ref{sec:purification}).
Various feedthroughs have been designed in order to bring the HVs and to extract the signal from the detectors, the environmental conditions of the gas and the video feed used for sparking monitoring.
Within this pressure vessel, the ThGEM detector stack is then installed for testing.

\subsubsection{Charge readout electronics}
\label{sec:electronics}
The readout cards used in the test setup are the ones previously used for MINOS \cite{7097530} and are the assembly of two cards. 
The Front-End Card (FEC) is a board originally built for the T2K experiment \cite{5321867}. 
It comprises four AFTER chips, and it can read out 288 channels. 
The AFTER chip has 72 channels.
The charge measurement range is selectable from 120 fC to 600 fC (120~fC and 600~fC used in our case depending on the signal size to limit saturation).
The peaking time can be set from 116~ns to 1912~ns, and was set to 1912~ns in here.
The maximum sampling rate of the AFTER chip is 100~MHz and is set to 3.33~MHz here leading to a time-bin width of 300~ns.
The depth of the switched capacitor array is 511 time-bins.
The Feminos is a small-size digital board designed to read out one FEC equipped with AFTER chips. 
The Feminos is based on a commercial FPGA module, the Mars MX2 from Enclustra, housed on a custom-made carrier card. 
The FPGA (Xilinx Spartan 6) implements all the necessary logic to configure and read out a FEC. 
It contains an embedded MicroBlaze processor for communication with a DAQ PC over Gigabit Ethernet.
Tests show that the maximum sustained data throughput of a Feminos is $\sim$~110~MByte/s.

The signal losses caused by the complete chain of electronics need to be measured.
Indeed, in the case of our measurements, the energy of the $\upalpha$-particle is precisely known.
However, for the use of the ThGEM with particles of unknown energy, the signal loss needs to be quantified.
For that purpose, the response of the electronics has been studied. 

\begin{figure}[htbp]
\centering
\includegraphics[width=0.55\textwidth, trim={0cm 0cm 0cm 1cm},clip]{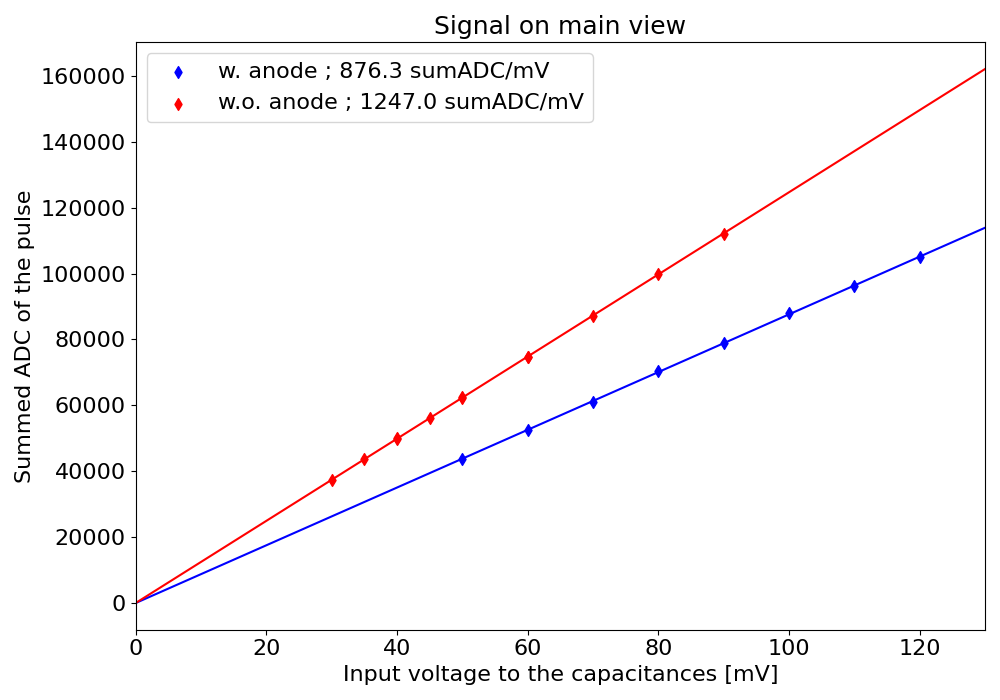}
\caption{Response of the electronics to an injected pulse, showing the effect of using versus not using the anode, and the anode impact on charge loss. The summed ADC is the sum of the maximum values from the waveforms in the different channels.\label{fig:electronics_response}}
\end{figure}

Estimating the impact of the electronics chain requires comparing signals injected through the anode and directly onto the electronics boards.
For each run at a different input voltage, the distribution of the peak value is fitted. The obtained results are displayed on Figure~\ref{fig:electronics_response}. 
We can see that the response of the electronics is linear with the input voltage, therefore the pulsed charge, in both cases. 
We can also see that the apparent charge is smaller when pulsing through the anode, as expected. 
The ratio between with and without the anode is of $\sim$~70\%.

The main conclusion of these studies is that the difference observed between the value of the charge measured with the anode alone and the charge produced by the radioactive source can be partially explained by the loss of charge in the electronics chain. 
This 70\% factor is one of the effects leading to the overall ADC to charge factor used in the gain estimation in our case since the $\upalpha$-particle energy is well-known.
These results clearly highlight the fact that an absolute calibration of the electronics response is crucial when detecting particles of unknown energy.
A loss of signal of at least 30~\% has been measured with the electronics used here.
Possibly, larger factors can be reached if longer cables are necessary between the detector and the electronics or additional capacitances are at play.
Failure to do a proper calibration would lead to an incorrect estimation of the gain of the ThGEM for instance or even to an improper measurement of the incident particle energy.

\subsubsection{Purification of the gas}
\label{sec:purification}
The presence of even trace amounts of contaminants in the gas can act as a quencher.
They theoretically enable a larger stability of the electrostatic detectors and can be purposely added to the operating gas (e.g. CF$_4$ and C$_4$H$_{10}$ for the Micromegas used in the TPCs of the near detector of the T2K experiment \cite{ATTIE2023168248}).
However, for some physics cases, the highest purity is required.
Indeed, the presence of quenchers would lead to the reduction of the ionisation signal during the drift and would limit the efficiency of the electron collection in the gas.
The argon gas used to fill the pressurised vessel is Argon Bip (6.0 quality, impurities lower than 1 ppm)  provided by the Air Products company(footnote : https://www.airproducts.fr/-/media/files/fr/250/250-17-062-fr-gaz-uhp-experis-argon.pdf).

\begin{figure}[htbp]
\centering
\includegraphics[width=0.6\textwidth]{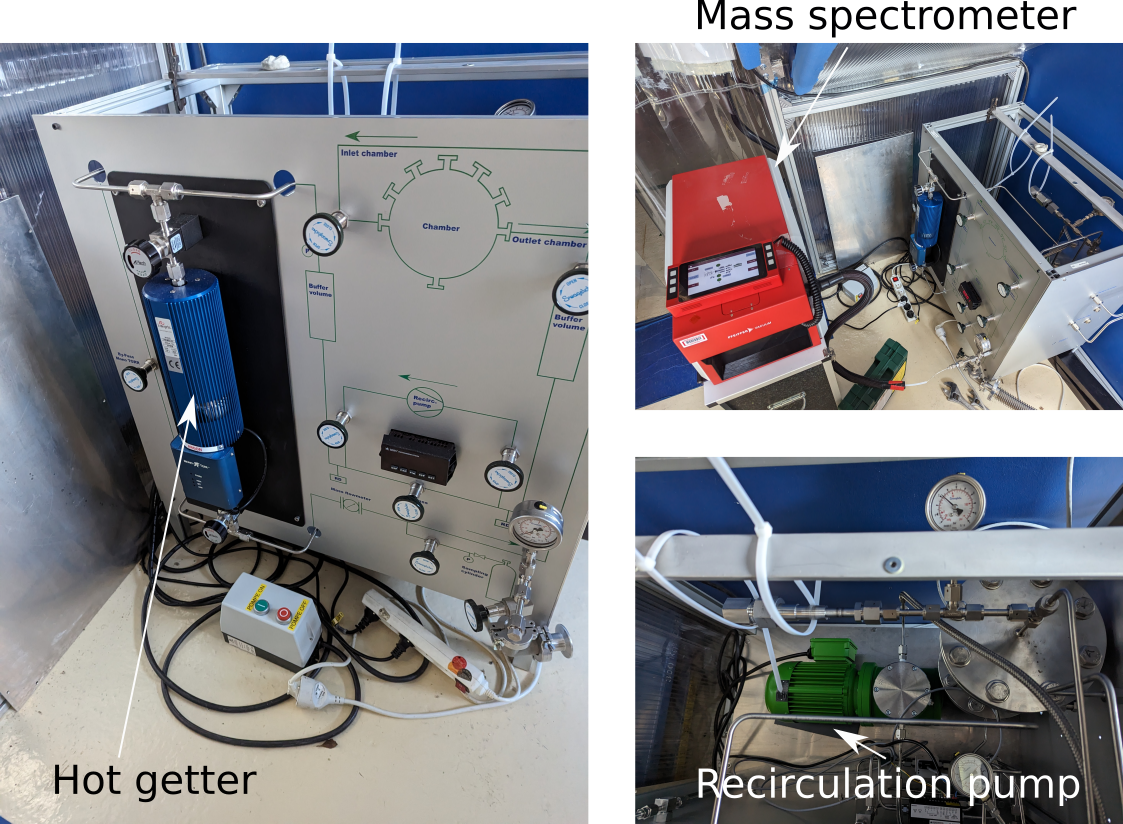}
\caption{Pictures of the purification system developed for the ThGEMs characterisation. The hot getter is visible on the left, while the gas recirculation pump and the mass spectrometer can be seen on the right.\label{fig:purification_system}}
\end{figure}

In noble gas dual-phase TPC, the impurities are actively removed by means of a purification system.
In order to reproduce these purity conditions, a dedicated purification system has been built as visible on Figure~\ref{fig:purification_system}.
This setup also allows the impact of the purification of the Argon on the performance of the ThGEMs to be quantified.
It encompasses a recirculation pump allowing a flow of several L/min leading to the exchange of the entire volume of the vessel in a few hours.
The purification of the gas is performed by a hot getter Entegris PS3-MT3-R.
It employs a cartridge containing a mixture of zirconium and titanium. 
When operated at high temperature, the getter absorbs impurities by forming irreversible bonds.
The purity levels are expected to reach below the ppb level in H$_2$O, O$_2$, CO, CO$_2$, H$_2$, CH$_2$ and N$_2$. 
In order to maximise the purification efficiency of the system, leaks have to be thoroughly investigated.
A mass spectrometer was available to us, but turned out not to possess the required sensitivity to estimate the level of the purification while providing us only with an upper limit of the contamination in the gas.
Due to the target levels of purity, external measurements of the purity were attempted in order to characterise the purification system itself.
The results of this measurement were inconsistent with our expectations, most likely resulting from a contamination of the gas sample.
We will assume here that the purification is effective and consider the specifications of the hot getter as our purity level.
Since no leak could be observed after several leak-checking campaigns and the relatively small volume of our setup compared to the recirculation speed, this is a valid assumption.

\section{Gain Measurements and main results}
\label{sec:gain}

\subsection{Gain measurement methodology}
\begin{figure}[h]
\centering
\includegraphics[width=0.45\textwidth]{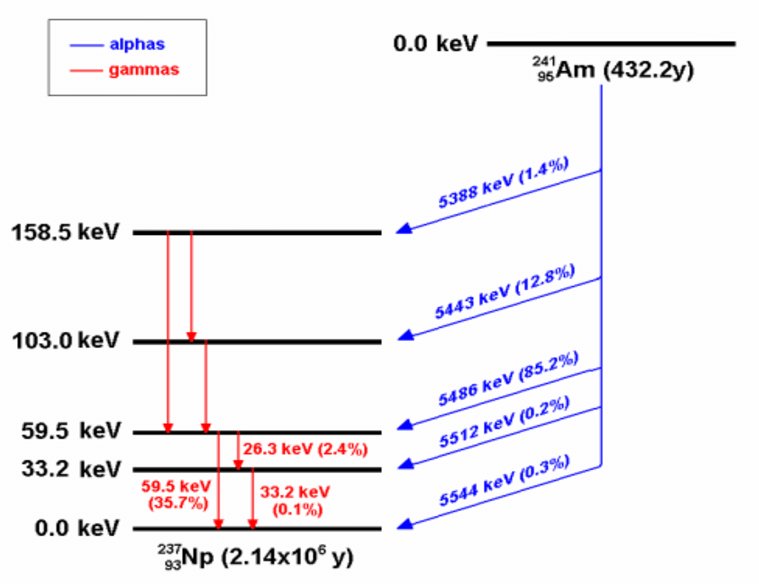}
\caption{Decay scheme of $^{241}$Am. Several $\upalpha$-decay lines can be observed. Their energies are however too close to be resolved by the ThGEM. Therefore, a value of 5.5~MeV will be assumed in the following.
\label{fig:am241}}
\end{figure}

In order to estimate the performance of the ThGEMs, the total gain needs to be measured.
A dedicated measurement using an Am$^{241}$ alpha-source is performed on each ThGEM.
The decay scheme of $^{241}$Am is displayed in Figure~\ref{fig:am241}.
The principle is to perform a HV scan, therefore varying the response of the detector due to a variation in the amplification field, while keeping a constant incoming energy deposit, at $\sim$~5.5~MeV corresponding to a primary charge of 33.2~fC or $\sim$~200k~e$^-$.
These measurements allow the different geometries and designs of ThGEMs to be compared to one another.

\subsection{Calculation of the gain of the ThGEM}
As visible in equation~\ref{eq:total_gain}, the total gain of the system is the convolution of several effects.
\begin{equation}
\label{eq:total_gain}
\begin{aligned}
G_{total} = \dfrac{Q_{measured}}{Q_{initial}} = \tau \times \epsilon_{collection}  \times G_{ThGEM} 
\end{aligned}
\end{equation}

$\tau$ is the overall transparency of the system, which will be described in detail in section~\ref{sec:transparency}.

\begin{figure}[h]
\centering
\includegraphics[width=0.6\textwidth]{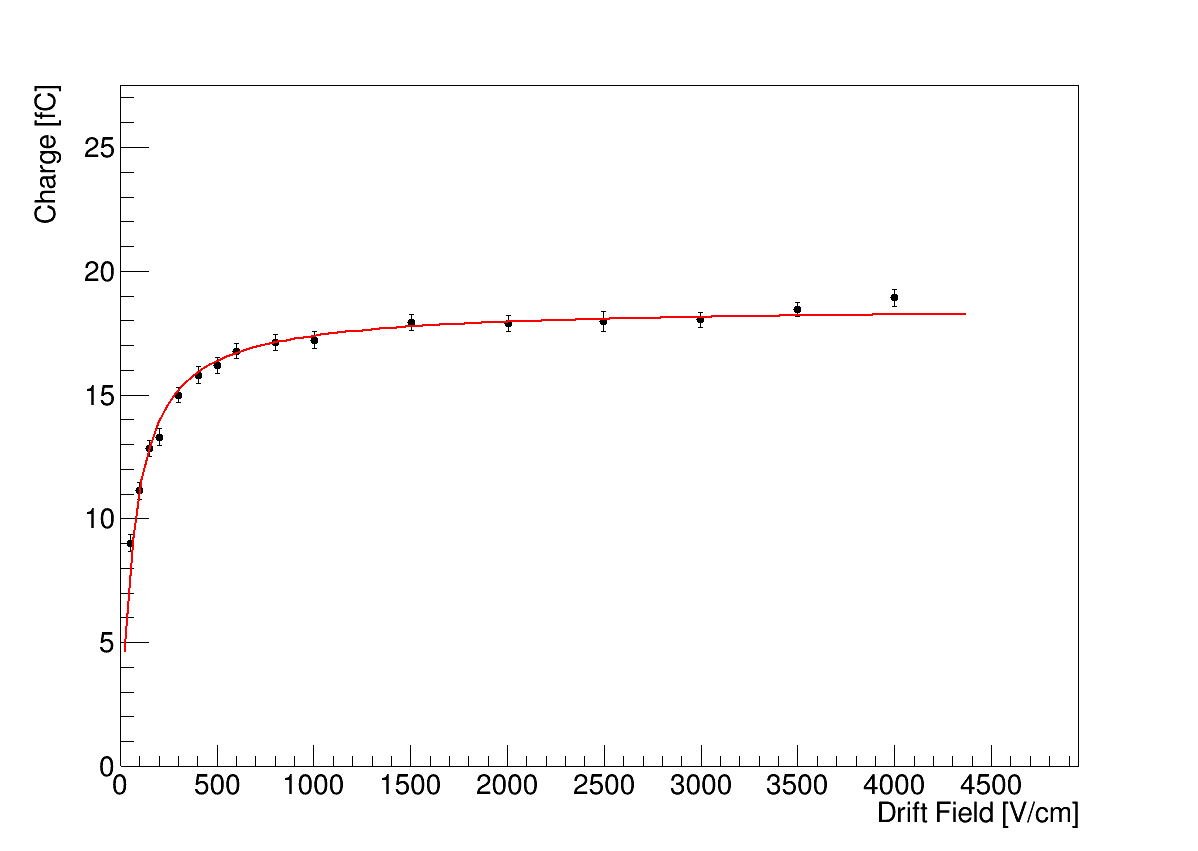}
\caption{Charge collected onto the 10$\times$10~cm$^2$ anode and read out by the electronics with respect to the drift electric field. The data follow Birks' law and a value of 16.2~fC at 500~V/cm can be extracted.
\label{fig:charge_collection}}
\end{figure}

$\epsilon_{collection}$ corresponds to the efficiency with which the electrons initially produced are collected by the anode and the corresponding charge readout by the electronics. This was measured in our test setup by removing the ThGEM and measuring the collected charge deposited by the $^{241}$Am source with respect of the drift electric field (see Figure~\ref{fig:charge_collection}). At the 500~V/cm drift field used in our setup for ThGEM gain measurements, 16.2~fC is collected and read out by the electronics, leading to $\epsilon_{collection}= 0.488$.
Within this factor of 48.8\%, several effects are at play.
Among them is the charge loss due to the electronics chain, described in section~\ref{sec:electronics}.

Finally, $G_{ThGEM}$ is the value of interest in our studies, corresponding to the absolute gain of the ThGEM stack.
As can be seen in \cite{Cantini_2015}, the effective gain of a ThGEM is proportional to:
\begin{equation}
\label{eq:gain}
\begin{aligned}
\text{G}_{\text{ThGEM}}(\text{E}, \uprho,t) \propto \text{e}^{\upalpha(\uprho,\text{E})x} \times \text{C (t)}
\end{aligned}
\end{equation}

where $\upalpha$ is the first Townsend ionisation coefficient for the amplification field E and density $\uprho$, x is the ThGEM thickness (1~mm in this case).
C (t) represents the time variation of the gain, which is negligible here once the charging-up of the detector is completed (see section \ref{sec:charging_up}).

The first Townsend ionisation coefficient, which corresponds to the number of electron-ion pairs produced by an electron per unit drift length, can be approximated by:
\begin{equation}
\label{eq:townsend}
\begin{aligned}
\upalpha(\uprho, \text{E}) = \text{A}\uprho \text{e}^{-\dfrac{\text{B}\uprho}{\text{E}}} 
\end{aligned}
\end{equation}

For reference, in the liquid argon conditions (87K and 0.980 bar), A$\uprho$ = (7339 $\pm$ 90) cm$^{-1}$  and B$\uprho$ = (183 $\pm$ 1) kV/cm \cite{Cantini_2015}.
These values will be used as input to fit the gain curves obtained in the results section \ref{sec:gain_results}.

The gain is extracted from the ADC values measured by the electronics and converted into a charge signal in fC.
The various parameters such as the electronics calibration values and the electrostatics transparency are then taken into account in order to extract the actual gain of the ThGEM.

\subsection{Data Analysis}
\subsubsection{Charge readout by the AFTER electronics chip}
\label{sec:signal_shape}

The charge readout is done by use of the AFTER chip.
The charge deposit is assumed to be instantaneous.
It translates into the following expression for the current:
\begin{equation}
I_{in}(t)=Q_{anode} \times \delta(t)
\end{equation}

where $I_{in}(t)$ is the current entering the AFTER chip, $Q_{anode}$ is the charge collected by the anode and $\delta(t)$ is the Dirac distribution.
Thus, the ADC response we get is the response of the AFTER chip to a Dirac pulse.
The response of the AFTER chip has been thoroughly extracted in other studies \cite{Ambrosi:2023smx} and a full waveform fit would constitute an interesting expansion of the work performed here.
Due to the charge collection principle, the total charge collected here is considered to be the maximal ADC value of the recorder Waveform converted into a charge in fC by way of the dynamic range of the electronics.
In our case, we consider the greatest ADC possibly read out by the electronics, which has a value of 4096~ADC.
Then, depending on the expected amplitude of the signal, and to avoid saturation, runs with a dynamic range of 120~fC and 600~fC have been recorded.
The appropriate value, i.e. corrected for the attenuation discussed in section~\ref{sec:electronics} is then input in the gain measurement.

\subsubsection{Event and waveform analysis}
\begin{figure}[h]
\centering
\includegraphics[width=0.6\textwidth]{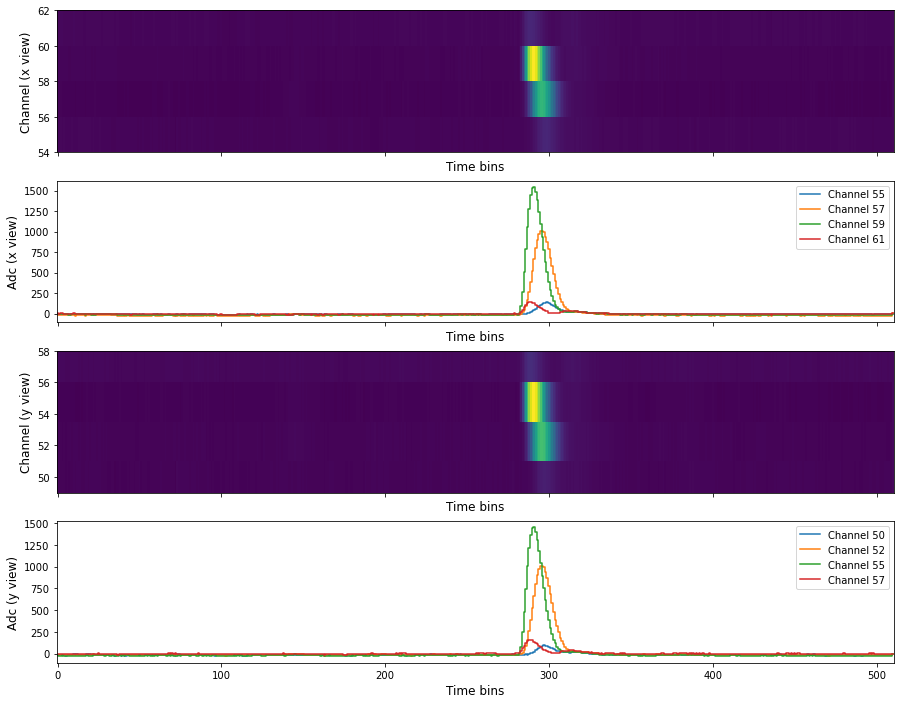}
\caption{Typical track recorded in the setup. The signal is spread across several channels, each providing a waveform, and read on the two orthogonal views in x and y.
\label{fig:event_display}}
\end{figure}

Due to the design of the anode, an energy deposit is read out on several channels and on two orthogonal views providing a relatively precise two-dimensional location for the energy deposit.
A typical event recorded via the anode of a ThGEM system is therefore composed of several waveforms.
In our studies using an $^{241}$Am source, the expected mean free path of the $\upalpha$-particle in argon gas at 3.3~bar is around 1.2 cm.
This corresponds to approximately 4 channels in either readout views.
Figure \ref{fig:event_display} is an example of an event recorded in our setup.

\begin{figure}[h]
\centering
\includegraphics[width=0.7\textwidth]{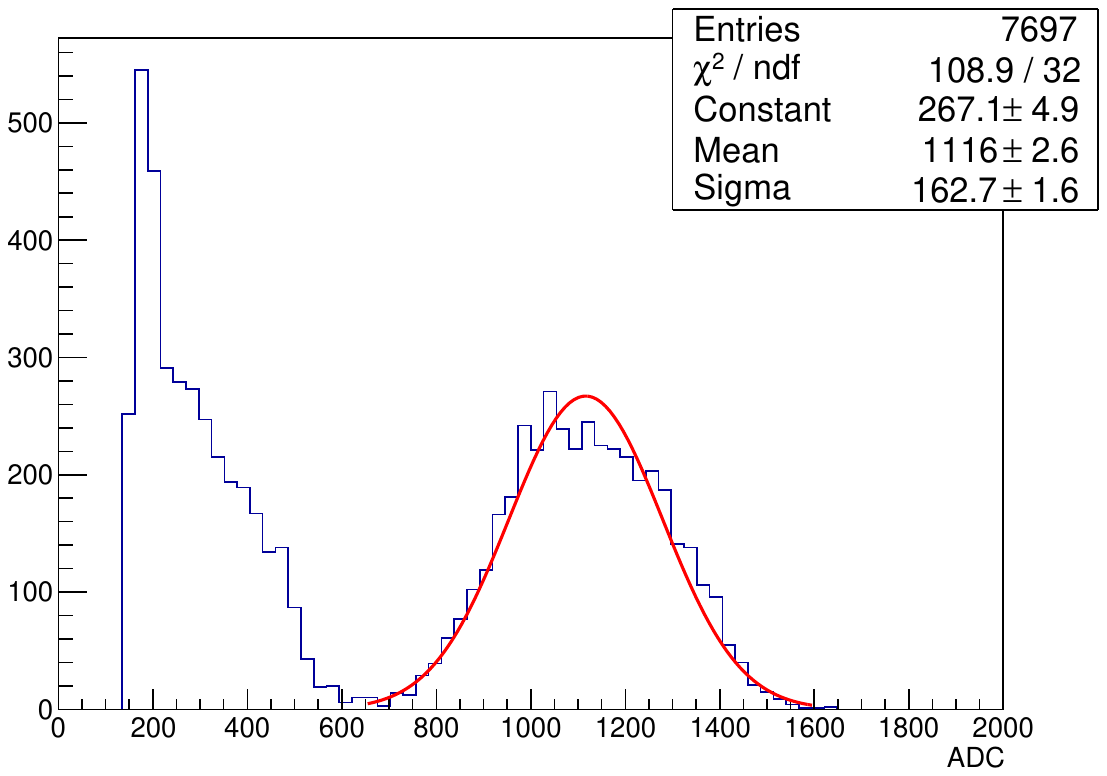}
\caption{Example of the Gaussian adjustment of the ADC distribution in a run. A lower cut is applied to the fit in order to remove noise and $\upgamma$ events. The mean value of the fit is converted into the ThGEM gain.}
\label{fig:gain_extraction}
\end{figure}

The method to estimate the performance, hence the gain, of a ThGEM in this work is to add the maximum ADC value in each of the channels triggered in the event.
This can then be converted into a deposited charge and the gain of the ThGEM can be calculated, knowing the initial energy of the $\upalpha$-particle. 
The estimation of the gain of a given electrostatic configuration is obtained by adjusting a Gaussian distribution to the ADC values collected for every event in a given run.
An example is displayed in Figure \ref{fig:gain_extraction}.
It can be seen that two populations are observed in the data.
The lower energy distribution is a superimposition of the electronics noise produced by the acquisition chain and the $\upgamma$-radiation background in the pressurised vessel.
The higher energy contribution comes from the $\upalpha$-radiation events produced by the $^{241}$Am source.

\section{Main Results}
\label{sec:results}

\subsection{Measurement of the charging-up effect}
\label{sec:charging_up}
\begin{figure}[h!]
\centering
\includegraphics[width=0.7 \textwidth]{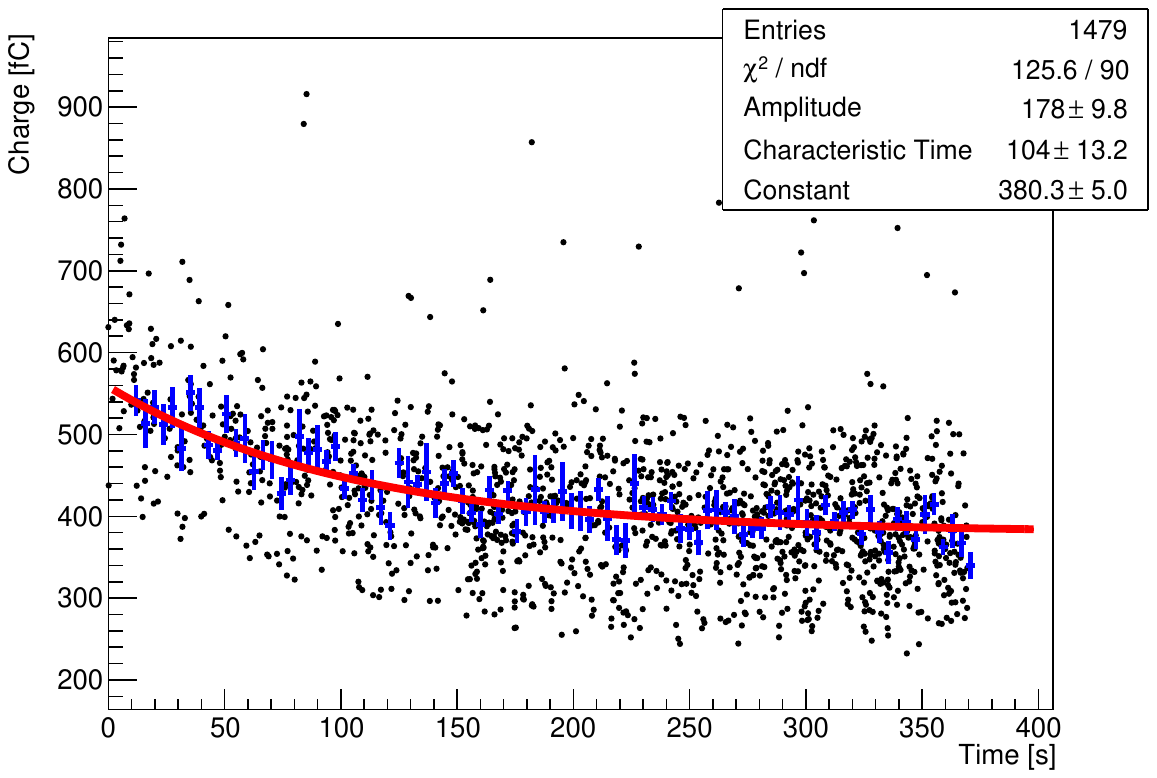}
\caption{Impact of the charging up on the gain at the beginning of a ThGEM operation. A clear reduction of the charge can be observed, by a factor of 1.45 in the example displayed here. The asymptote corresponds to the measured charge once the charging-up is complete. \label{fig:charging-up}}
\end{figure}

Due to the geometry of the ThGEMs, the gain evolves from the operation start and decreases by a large factor until it reaches a stable value.
This comes from the fact that the walls of the holes are made of FR4, an insulating material.
Prior to the start of operation, the surface of the holes are free of electrical charges.
When the first electrons start to pass through the holes, charges start to populate the walls, therefore impacting the electric field lines in the holes.
This has the consequence to reduce the gain within the ThGEM as can be seen in Figure \ref{fig:charging-up}. 
The measured charge is progressively reduced, by the reduction factor displayed here, during the establishment of the charging-up effect from its initial value to a fixed value.

Once the charging-up is established, it remains constant independently of the operation of the ThGEM.
The characteristic time of the evolution of the gain has been found to depend on the dimension of the ThGEM, the operation time, the energy of the initial particles and their amount.
This does not have a negative impact on the operation of these detectors since only the time necessary to reach the fully charged-up state evolves.

It is worth mentioning that, while not necessary for the operation of ThGEM, the only solution that was found to remove it is to physically immerse the ThGEM in a liquid so that the charges can be removed from the holes.
Indeed, the ions get neutralised in the solution and are washed away.
It is however only a temporary solution since the charging-up effect will appear again as soon as charges pass through the holes again.

\subsection{Electron transparency}
\label{sec:transparency}
The ThGEM gain is the factor by which the electrons released in the drift volume and passing through the ThGEM holes are multiplied by avalanche. The fraction of primary electrons going through the ThGEM holes is the electron transparency $\tau$ and depends on the optical transparency fixed by the geometry of the ThGEM (thickness, size and density of holes) and on the fraction of the electric field streamlines passing through the holes. COMSOL~MultiPhysics\textregistered~ simulations show that the electric field streamlines are mostly shaped by the electric field ratio between the drift electric field and the ThGEM amplification electric field.

\begin{figure}[h!]
\centering
\includegraphics[width=\textwidth]{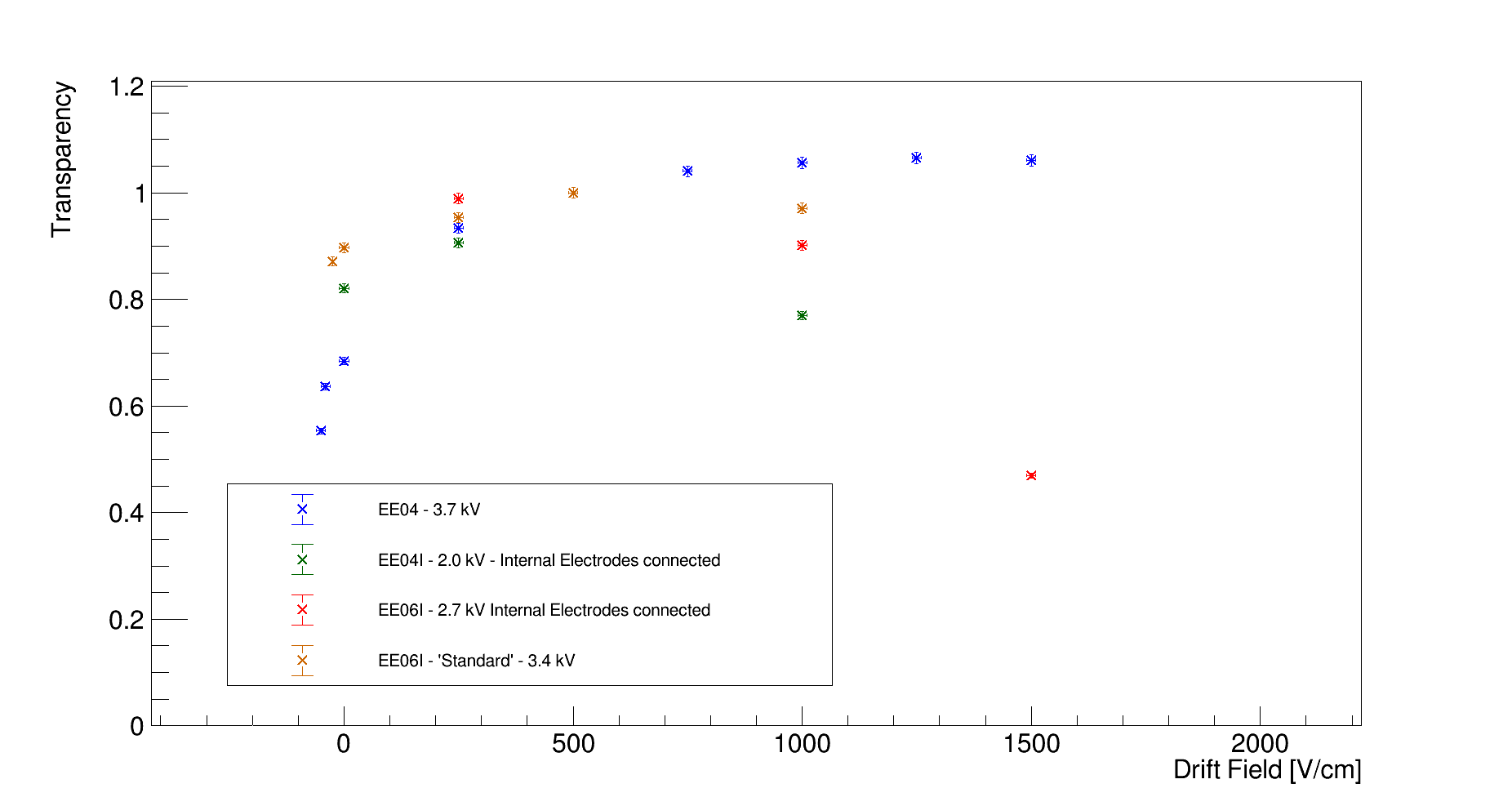}
\caption{Electronic transparency for EE04, EE04I and EE06I. Several behaviours can be observed depending on the electrostatic configuration. A plateau can be observed for EE04 while a decrease is visible at larger drift fields for the EE0xI detectors. The value at 500~V/cm is used for the normalisation. \label{fig:transparency}}
\end{figure}

Figure~\ref{fig:transparency} displays the transparency normalised to the value at 500~V/cm in each dataset, since it corresponds to the operational drift field for this work. It is a compromise in order enable the comparison of every design with one another.
The value of 500~V/cm has been chosen in order to maximise the transparency and is fixed for every design, which is confirmed by the results in Figure~\ref{fig:transparency}.
It is worth mentioning that a transparency of 100~\% is assumed for the gain measurements presented below.

It appears that the transparency has a strong dependence on the drift field value.
For the design with connected buried electrodes, it appears that the transparency reaches a plateau.
For every other configuration displayed here, the transparency reaches a value close to maximum around 500~V/cm before decreasing again when the drift field increases. 

\subsection{Impact of the gas density and the purification on the gain}
\begin{figure}[h!]
\centering
\includegraphics[width=0.7\textwidth]{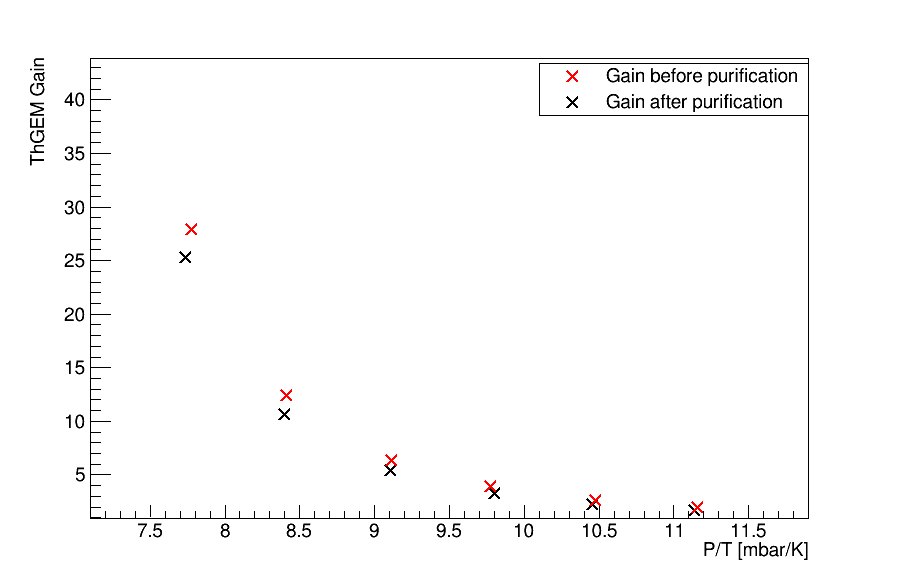}
\caption{Evolution of the EE04 ThGEM gain as a function of the density of the gas in the pressurised vessel for purified and unpurified argon. A lower gain is systematically observed for the operation in purified argon, here at 2700~V across the ThGEM. 
\label{fig:density}}
\end{figure}

The key aspects of the ThGEMs operations that are monitored here are the gains before and after purification.
The gas conditions have been defined here according to the applications in a liquid argon TPC.
This means that a significant fraction of the tests have been carried out with a pressure of 3.3 bar and in purified argon.
However, ThGEMs can operate at atmospheric pressure and up to large pressure values.
As can be seen in Figure~\ref{fig:density}, the gas density has an impact on the gain of the ThGEM.
It is illustrated here by the dependence of the gain with the pressure over temperature ratio of the gas.
For these measurements, temperatures around 20~${\degree}$C were measured.
When the pressure decreases, hence the density, a larger amount of primary electrons find their way to the ThGEM before recombining.
This leads to an increase in the gain.

Figure~\ref{fig:density} also highlights the impact of the purification of the gas.
While the unpurified dataset corresponds to gas provided by the Argon 6.0 bottle, the purification system has been running for a week for this purified dataset.
With the recirculation speed, it corresponds to an exchange of over ten times the volume of the pressurised vessel.
It can be seen that the purification of the argon tends to reduce the observed gain independently of the gas density.
This can be interpreted by the lack of impurities in the gas leading to a smaller amount of primary electrons produced.
These impurities appear to be more easily ionised by the interacting particle in the TPC than pure argon.
This effect also looks like it compensates for the lack of electronegative impurities, removed by the purification, which tend to reduce the number of primary electrons reaching the ThGEM.

\subsection{Gain measurements}
\label{sec:gain_results}
\begin{figure}[htbp]
\centering
\includegraphics[width=1.1\textwidth]{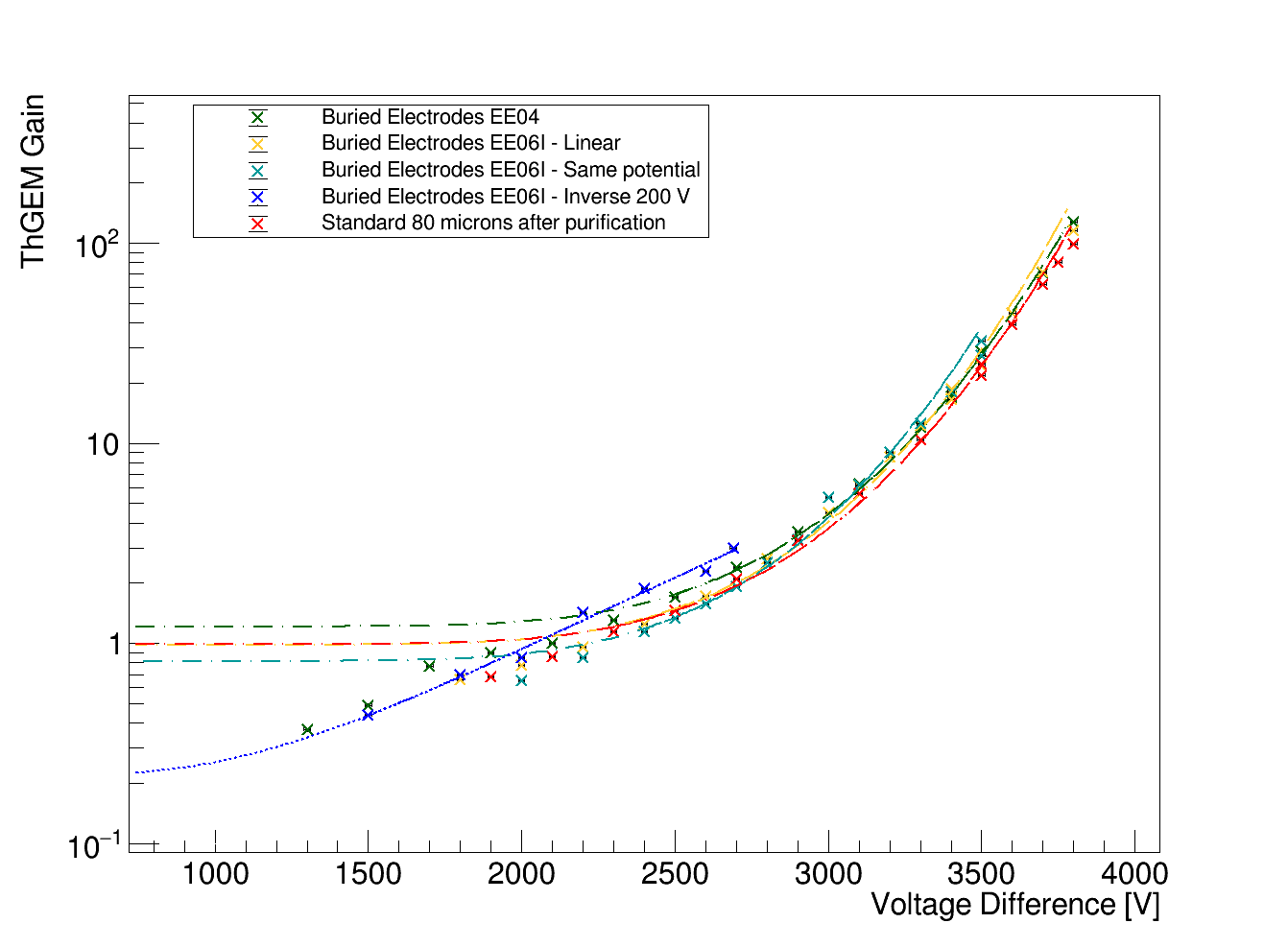}
\caption{Gain measured on the various designs of ThGEMs measured in 3.3 bar argon at room temperature. The line corresponds to the expected gain evolution of equation~\ref{eq:townsend} with the voltage difference within a ThGEM. The adjustment displayed for the inverse case does not follow equation~\ref{eq:townsend} with the expected Townsend coefficient and is therefore drawn in a different style for clarity.
\label{fig:gain_results}}
\end{figure}

Figure \ref{fig:gain_results} shows the results of various gain measurements that were carried out on the ThGEM detectors that were described in the previous sections. 
Measurements from the standard design (with 80~$\upmu$m rims) are compared with the EEx and EExI designs.
As can be seen, the evolution of the gains is comparable and follows the model of equation~\ref{eq:townsend} at high voltage difference.
At lower voltage difference, gain values below 1 are observed, which could mean that a minimal amplification field is necessary in order to allow the primary electrons to cross the ThGEM.

It should be noted that the evolution of the gain for the inverse configuration appears not to follow the gain evolution of the other configurations and detectors.
The potential difference used in Figure~\ref{fig:gain_results} is the value taken between the two outer electrodes.
The Townsend coefficients could not be used in this case for the fit of the gain evolution.  
This could be explained by a higher electric field for a given voltage difference compared to other configurations.
This would lead to higher gain values when the gain reaches values larger than 1, however a hard point at $\sim$2700~V across was reached.
A possible explanation could be intrinsic instabilities in the prototypes from the configuration that were not expected from the COMSOL\textregistered~ simulations.

The studies of the maximum voltage difference (across the external electrodes) aimed at getting the voltage difference for which the dead time was at maximum 30\%.
However, in the detectors that were studied, other issues are believed to have prevented such a systematic study.
Indeed, instabilities at relatively low potential values probably were caused by local defects of some holes.
This lead to the observation of carbonisation points and to electrical tripping of the HV channels of the power supply.
Therefore, the comparison of maximum voltages could not be carried out in this work.
It is however worth mentioning that operation voltage differences as large as 3800~V could be stably reached.

It is our belief that new detectors with the same design, and following the expectations of the COMSOL\textregistered~simulations, would show that the inverse configuration with independent buried electrodes would lead to larger gains thanks to larger maximum operation voltages.

\newpage

\section{Conclusion}
Operating gaseous detectors in pure noble gas is very challenging to obtain sufficiently high and stable charge amplification gains. For ThGEMs, one of the limitations comes from the high electric field at the edges of the holes which triggers sparks. COMSOL~MultiPhysics\textregistered~simulations of the electric field were done and show that adding two embedded internal electrodes in the ThGEM structure can be used to mitigate this. Several prototypes of these structures were produced, and their gain was measured in 6.0 quality pure argon at 3.3~bar and at room temperature. The gain in the ThGEM was derived from a complete set of measurements such as the electron transparency through the ThGEM, the collection efficiency of the segmented anode plane, and the calibration of the electronic readout chain coupled to the detector. Amplification gains in the ThGEMs up to 100 were reached, for amplification fields up to around 40~kV/cm, and fewer than 4 sparks per minute. Purifying the 6.0 quality argon gas circulating in the vessel with a hot getter leads to slightly lower gain values.

\acknowledgments
The authors would like to acknowledge the work performed by the student intern, A. Fritz, funded by IRFU, CEA, Université Paris-Saclay, Gif-sur-Yvette, France. We also acknowledge the support of CEA.

\bibliographystyle{JHEP}
\bibliography{biblio.bib}

\providecommand{\href}[2]{#2}\begingroup\raggedright\begin{thebibliography}{10}

\bibitem{GIOMATARIS199629}
Y.~Giomataris, P.~Rebourgeard, J.~Robert and G.~Charpak, \emph{Micromegas: a high-granularity position-sensitive gaseous detector for high particle-flux environments}, \href{https://doi.org/https://doi.org/10.1016/0168-9002(96)00175-1}{\emph{Nuclear Instruments and Methods in Physics Research Section A: Accelerators, Spectrometers, Detectors and Associated Equipment} {\bfseries 376} (1996) 29}.

\bibitem{Sauli:1997qp}
F.~Sauli, \emph{{GEM: A new concept for electron amplification in gas detectors}}, \href{https://doi.org/10.1016/S0168-9002(96)01172-2}{\emph{Nucl. Instrum. Meth. A} {\bfseries 386} (1997) 531}.

\bibitem{CHECHIK2004303}
R.~Chechik, A.~Breskin, C.~Shalem and D.~Mörmann, \emph{Thick gem-like hole multipliers: properties and possible applications}, \href{https://doi.org/https://doi.org/10.1016/j.nima.2004.07.138}{\emph{Nuclear Instruments and Methods in Physics Research Section A: Accelerators, Spectrometers, Detectors and Associated Equipment} {\bfseries 535} (2004) 303}.

\bibitem{Imaging}
L.~Gallego~Manzano, S.~Bassetto, N.~Beaupere, P.~Briend, T.~Carlier, M.~Chérel et~al., \emph{Xemis: A liquid xenon detector for medical imaging}, \href{https://doi.org/10.1016/j.nima.2014.11.040}{\emph{Nuclear Instruments and Methods in Physics Research Section A Accelerators Spectrometers Detectors and Associated Equipment} {\bfseries 787} (2015) }.

\bibitem{Cantini_2015}
C.~Cantini, L.~Epprecht, A.~Gendotti, S.~Horikawa, L.~Periale, S.~Murphy et~al., \emph{Performance study of the effective gain of the double phase liquid argon lem time projection chamber}, \href{https://doi.org/10.1088/1748-0221/10/03/P03017}{\emph{Journal of Instrumentation} {\bfseries 10} (2015) P03017}.

\bibitem{deoliveira}
R.~de~Oliveira for~the CERN MPT~Workshop, \emph{Thgem production (including new processes and new ideas)}, {\emph{https://indico.fnal.gov/event/23774/} (April,6 2020) }.

\bibitem{cotte:tel-02382815}
P.~Cotte, \emph{{Le projet WA105 : un prototype de chambre {\`a} projection temporelle {\`a} argon liquide diphasique utilisant des d{\'e}tecteurs LEMs}}, theses, {Universit{\'e} Paris Saclay (COmUE)}, Sept., 2019.

\bibitem{Cantini_2014}
C.~Cantini, L.~Epprecht, A.~Gendotti, S.~Horikawa, S.~Murphy, G.~Natterer et~al., \emph{Long-term operation of a double phase lar lem time projection chamber with a simplified anode and extraction-grid design}, \href{https://doi.org/10.1088/1748-0221/9/03/P03017}{\emph{Journal of Instrumentation} {\bfseries 9} (2014) P03017}.

\bibitem{grang2022}
P.~Granger, \emph{Studies of Time Projection Chambers using micro-pattern detectors for the DUNE experiment}, Ph.D. thesis, 2022.

\bibitem{comsol}
{COMSOL AB, Stockholm, Sweden}, \emph{{COMSOL Multiphysics\textregistered}},  v 6.3.
\newblock {www.comsol.com}.

\bibitem{7097530}
S.~Anvar, H.~Baba, H.~Bervas, D.~Besin, D.~Calvet, F.~Château et~al., \emph{The readout electronics and data acquisition system of the minos vertex tracker},  in \emph{2014 19th IEEE-NPSS Real Time Conference}, pp.~1--5, 2014, \href{https://doi.org/10.1109/RTC.2014.7097530}{DOI}.

\bibitem{5321867}
P.~Baron, D.~Besin, D.~Calvet, C.~Coquelet, X.~De~La~Broise, E.~Delagnes et~al., \emph{Architecture and implementation of the front-end electronics of the time projection chambers in the t2k experiment},  in \emph{2009 16th IEEE-NPSS Real Time Conference}, pp.~43--48, 2009, \href{https://doi.org/10.1109/RTC.2009.5321867}{DOI}.

\bibitem{ATTIE2023168248}
D.~Attié, O.~Ballester, M.~Batkiewicz-Kwasniak, P.~Billoir, A.~Blanchet, A.~Blondel et~al., \emph{Analysis of test beam data taken with a prototype of tpc with resistive micromegas for the t2k near detector upgrade}, \href{https://doi.org/https://doi.org/10.1016/j.nima.2023.168248}{\emph{Nuclear Instruments and Methods in Physics Research Section A: Accelerators, Spectrometers, Detectors and Associated Equipment} {\bfseries 1052} (2023) 168248}.

\bibitem{Ambrosi:2023smx}
L.~Ambrosi et~al., \emph{{Characterization of charge spreading and gain of encapsulated resistive Micromegas detectors for the upgrade of the T2K Near Detector Time Projection Chambers}}, \href{https://doi.org/10.1016/j.nima.2023.168534}{\emph{Nucl. Instrum. Meth. A} {\bfseries 1056} (2023) 168534} [\href{https://arxiv.org/abs/2303.04481}{{\ttfamily 2303.04481}}].

\end{thebibliography}\endgroup


\begin{thebibliography}{99}

\bibitem{a}
Author,
\emph{Title},
\emph{J. Abbrev.} {\bf vol} (year) pg.

\bibitem{b}
Author,
\emph{Title},
arxiv:1234.5678.

\bibitem{c}
Author,
\emph{Title},
Publisher (year).

\end{thebibliography}

\end{document}